\documentclass[reprint,aps,prb,10pt,twocolumn,superscriptaddress,showpacs,longbibliography,amsmath,amsfonts,amssymb,floatfix]{revtex4-1}

\usepackage[pdftex]{graphicx}
\usepackage{float} \usepackage{color}

\usepackage[pdftex,colorlinks=true]{hyperref}

\usepackage{mathtools} 

\DeclarePairedDelimiterX\braket[2]{\langle}{\rangle}{#1 \delimsize\vert #2}
\DeclarePairedDelimiterX\expval[3]{\langle}{\rangle}{#1 \delimsize\vert #2
  \delimsize\vert #3}

\newcommand{\vect}[1]{\mathbf{#1}}

 \newcommand{\bs}{{\bf{s}}}
 \newcommand{\bL}{{\bf{l}}}

\newcommand{\bea}{\begin{eqnarray}} \newcommand{\eea}{\end{eqnarray}}

\newcommand{\beq}{\begin{equation}} \newcommand{\eeq}{\end{equation}}

\def\sectionn#1{\noindent\underline{\it #1:}}

\begin{document}

\title{Dynamics and energy landscape of the jammed spin-liquid}

\author{Thomas Bilitewski} \affiliation{Max-Planck-Institut f\"{u}r Physik
  komplexer Systeme, N\"othnitzer Str.\ 38, 01187 Dresden, Germany}

\author{Mike E. Zhitomirsky} \affiliation{Universite Grenoble Alpes, CEA,
  INAC-Pheliqs, 38000 Grenoble, France}

\author{Roderich Moessner} \affiliation{Max-Planck-Institut f\"{u}r Physik
  komplexer Systeme, N\"othnitzer Str.\ 38, 01187 Dresden, Germany}

\begin{abstract}
  We study the low temperature static and dynamical properties of the classical
  bond-disordered antiferromagnetic Heisenberg model on the kagome lattice. This
  model has recently been shown to host a new type of spin liquid exhibiting an
  exponentially large number of {\it discrete} ground states. Surprisingly,
  despite the rigidity of the groundstates, we establish the vanishing of the
  corresponding spin stiffness. Locally, the low-lying eigenvectors of the
  Hessian appear to exhibit a fractal inverse participation ratio. Its spin
  dynamics resembles that of Coulomb Heisenberg spin liquids, but exhibits a new
  low-temperature dynamically arrested regime, which however gets squeezed out
  with increasing system size. We also probe the properties of the energy
  landscape underpinning this behaviour, and find energy barriers between
  distinct ground states vanishing with system size. In turn the local minima
  appear highly connected and the system tends to lose memory of its inital
  state in an accumulation of soft directions.
\end{abstract}

\maketitle

\section{Introduction}
Complex energy landscapes are of interest in a variety of fields, from
(combinatorial) optimisation problems \cite{Reidys2002,Franz2017} over the
physics of spin glasses
\cite{Anderson1975,Mezard1984,Mezard1986,Young1997,Charbonneau2014,Charbonneau2017,Pemartin2018},
jamming \cite{Charbonneau2017,Liu2010,Liu1998,Ohern2003} and amorphous materials
\cite{Berthier2011,Charbonneau2017,Behringer2018}, to the folding of biopolymers
\cite{Onuchic2000}, chemical reactions \cite{Heidrich1991} and the fitness
landscape of evolution \cite{Wright1932,Mustonen2009,Stadler2002,Hartl2014}.
Their phenomenology can be formulated in terms of the nature of these energy
landscapes, their geometric features, e.g. their ruggedness, the structure of
the minima and barriers between them, in terms of the dynamics of systems
evolving within them, and the relation between the static and dynamic
properties.

Here, we study these questions in a classical frustrated magnet with
bond-disorder which hosts a jammed spin liquid, jammed in the sense that in
groundstates the number of spin degrees of freedom is exactly balanced by the
number of independent constraints on the system, in analogy to the critical
point of the jamming transition in granular media \cite{Behringer2018,Liu2010}
at which motion is arrested by contacts between particles at the jamming
transition.

Finding energy minima of ``glassy'' systems is (often) NP hard
\cite{Barahona1982}. Here, an extensive number of exactly degenerate ground
states with a known minimal energy arises in the presence of disordered
couplings. This allows us to make a sharp distinction between meta-stable,
excited states and groundstates. This tends to be more difficult in disordered
systems when the true minimal energy is not known. It also allows a sharp
definition of energy barriers between different groundstates as their energy is
known a priori to be the same.

In geometrically frustrated magnets ordering is suppressed due to competing
interactions, which in classical systems leads to a large number of degenerate
ground states \cite{Anderson1956,Villain1979,Chalker1992,Moessner1998}. A
paradigmatic example of geometric frustration in this sense is the
nearest-neighbour Heisenberg antiferromagnet (HAFM) on the Kagome lattice
\cite{Chalker1992,Huse1992,Reimers1993,Moessner1998}, with a cooperative regime
extending from $T \sim 0.1 J$ down to $T \sim 0.001 J$ eventually terminated by
an order-by-disordered octupolar regime \cite{Zhitomirsky2008,Chern2013}.



Recently, it has been shown that there is an intimate connection between this
groundstate degeneracy of the Kagome HAFM and topological quantities via
generalized origami mappings in the case of anisotropc interactions
\cite{Roychowdhury2018a,Roychowdhury2018}.

Interestingly, weak bond disorder in the kagome HAFM does not produce a spin
glass, but rather defines a new type of spin liquid, dubbed a jammed spin-liquid
\cite{Bilitewski2017}.
In this case disorder removes all zero modes and prevents the entropic
order-by-disorder selection of coplanar states, and the ground state manifold
remains disordered down to the lowest temperatures. This motivates the current
study: We are seeking to understand in detail the properties of the groundstate
manifold and the resulting dynamics of the jammed spin liquid in the complex
disordered energy landscape.


We find the following phenomenology: The spin dynamics resembles that of other
$U(1)$ Coulomb Heisenberg spin liquids with exponentially decaying
spin-autocorrelation functions, and broad features in the dynamical structure
factor showing no indication of well defined quasi-particle excitations. At
extremely low temperatures, which vanish in the thermodynamic limit as $L^{-3}$,
the system is dynamically arrested and trapped close to a single ground state.
The low-lying part of the spectrum of the Hessian, describing nature of local
fluctuations around a given energy minimum, involves modes whose inverse
participation ratio (IPR, Eq.~\ref{eq:IPR}) is best fit by a fractal decay with
system size, $L^{-5/3}$. The energy barriers between different groundstates are
found to decrease with system size as $L^{-3}$. However, it appears that such
transitions between groundstates require delocalised changes of the whole spin
configuration, while local perturbations encounter significantly enhanced
energy-barriers. Finally, we find that successive transitions enable states to
explore a large part of the ground state manifold, completely loosing memory of
the initial state in an exponential fashion. Thus, in this disordered frustrated
magnet an energy landscape of discrete degenerate groundstates separated by
thermodynamically vanishing energy-barriers that appears to be (at least partly)
connected emerges.

The remainder of the manuscript is structured as follows: We introduce model and
the numerical procedures in section~\ref{sec:model}. We first discuss the
spin-stiffness of the ground states in section~\ref{sec:spin_stiffness}. Then we
explore the classical spin dynamics, including the spin autocorrelation and the
dynamical structure factor as well a transition to a dynamically arrested state,
in sec~\ref{sec:dynamics}. We then address the nature of the energy landscape,
first in terms of the statistical properties of their Hessian matriices in
section~\ref{sec:hessian}. We continue with a detailed study of the
groundstates, first their response to applied fields in sec.~\ref{sec:forcing},
and in terms of a random walk in the space of ground states in
sec~\ref{sec:random_walk}. We summarise our main findings and conclude in
sec.~\ref{sec:conclusions}.

\section{Model\label{sec:model}}
\begin{figure}
  \begin{minipage}[t]{0.99\columnwidth}
    \vspace{0pt}
    \includegraphics[width=.99\columnwidth]{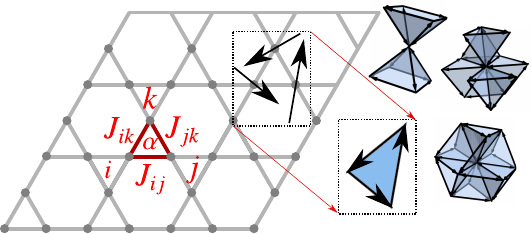}
  \end{minipage}
\caption{Illustration of finite-size kagome lattices with primitive
  lattice vectors $\vect{a}_1=(1,0)$ and $\vect{a}_2=(1,\sqrt{3})/2$. Classical $O(3)$ spins in
  triangle $\alpha$ at sites $i,j,k$ are coupled anti-ferromagnetically via $J_{ij}>0$.
  Groundstates are defined via local constraints $\vect{l}_{\alpha}=0$, such that  spins on every triangular plaquette form a scalene triangle in the spin space and
produce generically non-coplanar magnetic structures illustrated on the right-hand side.
  \label{fig:illustration_lattices}
}
\end{figure}
\sectionn{Hamiltonian} We consider the classical nearest neighbour Heisenberg
model \bea \mathcal{H}&=&\sum_{\langle ij\rangle} J_{ij} \, \bs_i\cdot\bs_j \
,
\label{eq:KAFM}
\eea with disordered anti-ferrromagnetic couplings $J_{ij}>0$ between $O(3)$
spins ($\left| \vect{s}_i \right|=1$) at site $i,j$ on the Kagome lattice as
illustrated in Fig.~\ref{fig:illustration_lattices}.

The Hamiltonian can be rewritten as a sum of squares
\begin{equation}
  \mathcal{H}=\frac{1}{2} \sum_\alpha \bL_\alpha^2  +\textrm{const},\quad \text{with} \quad \bL_\alpha=\sum_{i\in\alpha}\gamma_{i\alpha}\bs_i \, ,
  \label{eq:H_general}
\end{equation}
where in every triangle $\alpha$ formed by sites $ijk$ we defined
$\gamma_{i\alpha}=\sqrt{{J_{ij}J_{ik}}/{J_{jk}}}$. Both forms are completely
equivalent if $J_{ij}>0$ allowing to define $\gamma_{i \alpha}$.

We will mainly work with the second form, and restrict the model further by
requiring $\gamma_{i\vartriangle} = \gamma_{i\triangledown}$, which corresponds
to some short-range correlations of the bond-couplings. This is done mainly to
reduce finite-size effects, in particular, for the ground state energy which is
$E_{\rm g.s.}= 0$ (ignoring the constant term) once all constraints are
satisfied. The groundstates of these models exhibit different spin-spin correlations,
in the first case being exponentially decaying, and in the second algebraically
decaying. However, the finite temperature and dynamical behaviour appears qualitatively and
quantitatively similar. This modification is not expected to change the results
of this study qualitatively which has been explicitly confirmed for the dynamics. 

\sectionn{Groundstate manifold} From Eq.~\ref{eq:H_general} states that satisfy
$l_{\alpha}=0$ on all triangles are seen to be ground states. This can be
interpreted as the sum of spins with different length scaling factors
$\gamma_{i\alpha}$ vanishing, i.e. forming a closed triangle in spin space as
shown in Fig.~\ref{fig:illustration_lattices}. The resulting spin configuration
of a single triangle is coplanar, but generally non-collinar, while on the full
lattice it becomes non-coplanar as well. It may be visualised as a
three-dimensional structure with scalene triangles as faces, see
Fig.~\ref{fig:illustration_lattices}.

The fact that all constraints on the kagome lattice can be satisfied
simultaneously is non-trivial. The resulting set of groundstates of the jammed
spin liquid \cite{Bilitewski2017} includes exponentially many exactly degenerate
non-coplanar groundstates in presence of disorder (up to a critical disorder
strength), which are rigid without any zero-modes besides global rotations.
In particular, they are not connected to the coplanar groundstates, which are
known to determine the low-temperature properties of the non-disordered model
\cite{Chalker1992,Huse1992,Reimers1993,Moessner1998,Zhitomirsky2008}, and have
an extensive number of zero-modes \cite{Chalker1992,Huse1992}; rather they form
a disconnected discrete set, instead of a continuous connected manifold
\cite{Moessner1998}.


\sectionn{Dynamics} The semi-classical spin dynamics, describing precession of
spins around their local exchange fields, is given by the Landau-Lifshitz
equation \cite{Landau1975},
\begin{equation}
  \frac{d \vect{s}_i(t)}{dt} = - \vect{s}_i(t)  \times \left( \sum_{j} J_{ij} \vect{s}_j(t) \right)
  \label{eq:bloch_eq}
\end{equation}
which conserves the total energy $E$, magnetisation $\vect{M}$ as well as the
spin norm.

\sectionn{Global Symmetries and equivalence classes of states} The Hamiltonians in
Eq.~\ref{eq:KAFM} and \ref{eq:H_general} possess a global $O(3)$ symmetry. The
invariance of the energy under these rotations results in 3 global-zero modes of
the groundstates and the conservation of the total magnetisation under the
dynamics.
In defining distinct states it is necessary to take these symmetries into account.
Formally, one may use equivalence classes defining distinct states as
spin-configurations modulus the $O(3)$ symmetry. One may transform any spin configuration to
a representative of the equivalence class, e.g. by rotating the spins such that ${s}_1$ points in a
fixed direction by using a global rotation of all spins and $\vect{s}_2$ lies in
a fixed plane by rotating all spins around $\vect{s}_1$, and compare spin
configurations after rotating into this fixed frame. Alternatively, the
gram-matrix $g_{ij} = \vect{s}_i \cdot \vect{s}_j$ uniquely characterises distinct
equivalence classes accounting for the rotational symmetry automatically. In
practice, we choose to work with explicit representatives by rotating into a
fixed frame for this study.

\sectionn{Details on the numerics} We perform both Monte-Carlo simulations to
obtain finite temperature spin configurations and explicit energy minimisation
to obtain ground state spin configurations. Both are combined with molecular
dynamics simulations
\cite{Moessner1998,Moessner1998a,Conlon2009,Robert2008,Taillefumier2014}.

For the ground state simulations states are converged to an energy of
$E<10^{-14}$, or until the norm of the energy-gradient is smaller than $10^{-8}$
(in case we end up in a local minimum).

The Monte-Carlo simulations are performed using heat-bath updates combined with
micro-canonical overrelaxation updates. From these we obtain samples from the
Boltzmann distribution $\sim \exp[-\beta \mathcal{H}]$ at inverse temperature
$\beta$.

Taking the samples obtained via Monte-Carlo as initial conditions, the equations
of motion Eqs.~\ref{eq:bloch_eq} are integrated numerically, and quantities of
interest computed from the time-evolved spin configuration. Thus, the ensemble
averaged is approximated by an average over different initial states $\left<
  \vect{s}_i(t)\cdot \vect{s}_j(0) \right> \approx 1/N_{\textrm{states}}
\sum_{\textrm{states}} \vect{s}_i(t)\cdot \vect{s}_j(0)$. Time integration is
performed using a fourth-order Runge-Kutta algorithm with adaptive time step
size such that the error on the conserved energy, spin-length and magnetisation
remains below $10^{-6}$ per spin.

We study systems up to linear system size $L=24$ (corresponding to $N_s=1728$
spins) with explicit energy-minimisation, and systems up to $L=96$ ($N_s=27648$)
and temperatures $\beta=1,\cdots,10000$ with MC.

Throughout we work in dimensionless units with the lattice spacing $a=1$. We
choose the couplings $\gamma_{i}=1 +\delta_{i}$ with $\delta_{i}$ uniformly in
$[1-\delta,1+\delta]$ for disorder strength $\delta$. We also restrict to
$\delta=0.3$ in this work, but have checked that results are qualitatively the
same within the jammed spin liquid regime $\delta < 1/3$. Results are averaged
over 100 disorder realisations for the groundstate simulations, and over a 1000
disorder realisations for the MC simulations.

\section{Spin stiffness\label{sec:spin_stiffness}}

\subsection{Analytical derivation}
The spin stiffness is defined via the energy response to a twist, i.e. via
comparing the energy of states obtained with periodic boundary conditions (PBC),
and those with twisted boundary conditions along one of the lattice directions.
Specifically, we take $\vect{S}_{i+Le_x}= R(\theta, \vect{e}_{\theta})
\vect{S}_i$ with a rotation matrix $R$ depending on the twist angle $\theta$ and
the rotation axis $\vect{e}_{\theta}$. The energy difference between PBC and
twisted BC follows as the minimum over all possible orientations of the rotation
axis $\vect{e}_{\theta}$.

The vanishing spin stiffness in the jammed spin liquid regime can be derived
from considerations of the constraints defining the set of ground states,
together with the implicit function theorem. Specifically, we have that for
groundstates $\vect{l}_{\alpha}=0$ on all triangles $\alpha$. Imposing twisted
boundary conditions amounts to changing the energy function in the border
triangles in the following way:
\begin{equation}
  (\gamma_a \vect{S}_a + \gamma_b \vect{S}_b + \gamma_{c} \vect{S}_{c})^2 \rightarrow  (\gamma_a \vect{S}_a + \gamma_b \vect{S}_b + R(\theta,e_{\theta}) \gamma_{c} \vect{S}_{c})^2
\end{equation}
Thus, the zero-energy groundstates can still be written as a sum of squares, and
we have a mapping
\begin{equation}
  \label{eq:inv_function}
  \begin{split}
    G : \quad &\mathbb{R} \times \mathbb{R}^{3 N_{\mathrm{s}}}\rightarrow \mathbb{R}^{3N_{\mathrm{s}}} \\
    & \theta \times \{ \vect{S}_i\} \mapsto
    \begin{cases}
      \vect{S}^2_i-1  & i \in 1,\dots,N_S \\
      \bL_{\alpha}(\theta) & \alpha \in 1, \dots, 2 N_{\mathrm{s}}/3
    \end{cases}
  \end{split}
\end{equation}
where now $\bL_{\alpha}$ depends on the twisting angle $\theta$. The ground
state configurations for PBC then correspond to the preimage of the zero-vector,
e.g. $\{ \vect{S}^{\mathrm{gs}}_i\} = G^{-1}(\vect{0})$.

Given a ground state for PBC, e.g. a point $\{\theta_0=0,\{\vect{S}_i\}\}$ such
that $G(\{\theta_0,\{\vect{S}_i\}\})=\vect{0}$, the implicit function theorem
guarantees that the ground state is given by a differentiable function of the
twist angle $\theta$ in an open neighbourhood of $\theta_o$ if the Jacobian
$\left[ \frac{\partial G_i}{ \partial S_{j d}} \right]$ is invertible. Here
$j=1,\dots,N_{\mathrm{s}}$ is the site index and $d=x,y,z$ is the index of the
spatial dimension.

Since $R(0,e_{\theta})=1$ and our previous work already established the
non-vanishing of the Jacobian determinant for JSL ground states
\cite{Bilitewski2017}, we conclude that we can continue these states over a
finite range of twisting angles $\theta$ with exactly vanishing energy, thus
establishing the vanishing of the spin stiffness in the jammed spin liquid.

\subsection{Numerical Results}
\begin{figure}
  \begin{minipage}[t]{0.99\columnwidth}
    \vspace{0pt}
    \includegraphics[width=.99\columnwidth]{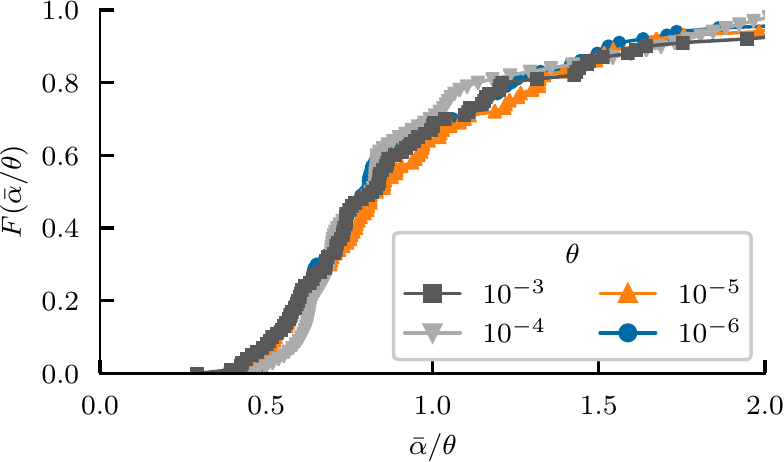}
  \end{minipage}
  \caption{Cumulative distribution $F(x) = \int_{0}^x p(y) dy$ of the scaled mean angle $\bar{\alpha}/\theta$ between states found for periodic
    boundary conditions and the one after applying a twist for twist angles
    $\theta=10^{-3}$ down to $\theta=10^{-6}$  obtained from 100 different ground states and disorder
    configurations.
  \label{fig:twisted}
}
\end{figure}
Beyond this proof of vanishing spin stiffness we consider the response of the
system to a twist in more detail numerically. To do so we obtain a ground state
for PBC via energy minimisation starting from a random initial configuration,
then apply the twist and start the energy minimisation from the previously found
state.

As a first check on the numerics, and to ensure that the $\theta$-range over
which states can be continued is (sufficiently) large in practice, we compute
the average rotation angle between the spin configuration found for twisted BC
and the one for PBC defined as
\begin{equation}
  \bar{\alpha}=1/N_s \sum_i \arccos(\vect{S}^{PBC}_i \cdot \vect{S}^{TBC}_i(\theta))\, .
\end{equation}
This should stay small and be linear in $\theta$, such that the twisted state
remains close to the initial state, and the energy difference actually is a
measure of the spin stiffness of that state.

We show the cumulative distribution function of the scaled twist angle
$\bar{\alpha}/\theta$ obtained from 100 different disorder realisations and
states on a $L=12$ system in Fig.~\ref{fig:twisted} for a range of twisting
angles $\theta = 10^{-6}$ up to $10^{-3}$. The collapse of the data confirms the
expected linear scaling of the response to the twist which remains on the
natural order of $\theta$.

Importantly, we find the energy difference to vanish within the numerical accuracy for
sufficiently small twist angles, specifically we checked it for twisting angles
$\theta=10^{-6}$ up to $\theta=10^{-3}$ for different initial states and
disorder realisations and different system sizes $L=6,12$, i.e. we confirm the
vanishing of the spin stiffness also numerically.
\section{Spin Dynamics\label{sec:dynamics}}

\subsection{Spin Autocorrelation}
The simplest indicator of the nature of spin dynamics is the spin
autocorrelation function defined as
\begin{equation}
  A(t) = \frac{1}{N}\sum_i \left< \vect{s}_i(t) \cdot \vect{s}_i(0)\right> \, ,
\end{equation}
which may be interpreted as the overlap between the initial and time-evolved
state.

\begin{figure}
  \includegraphics[width=.99\columnwidth]{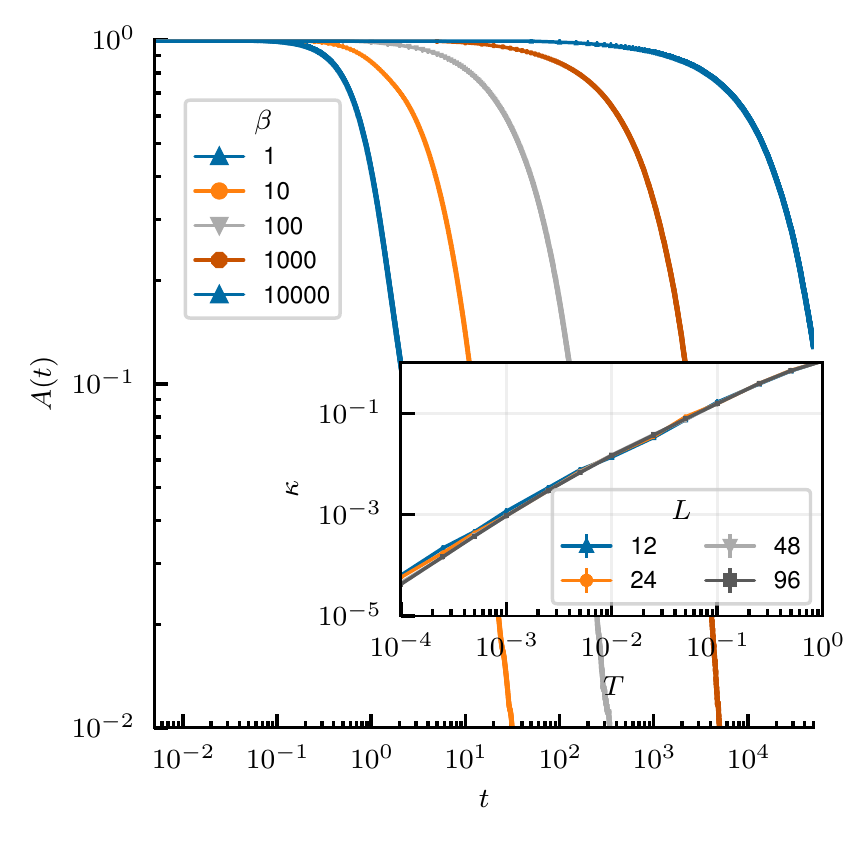}
  \includegraphics[width=.99\columnwidth]{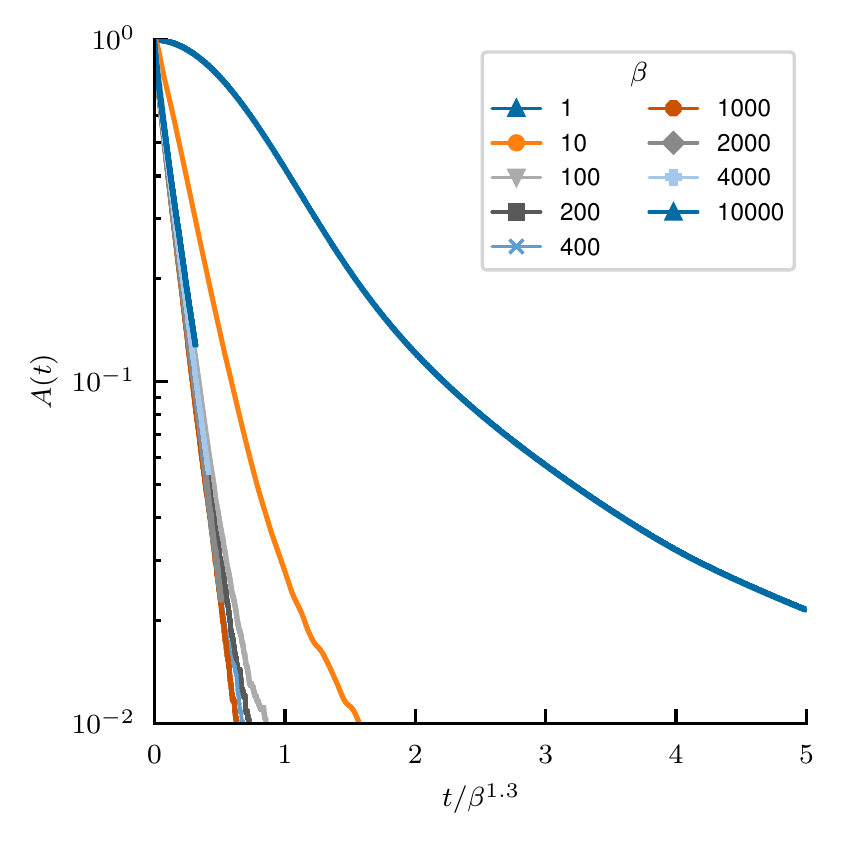}
\caption{Main: Spin autocorrelation on a $L=96$ system at temperatures
  $\beta=1,\cdots,10000$ as indicated in the legend
  Top on log-log scale versus time $t$, bottom on log-linear scale versus scaled time $t/\beta^{1.3}$.
  Inset: Extracted exponential decay constant $A(t) \sim e^{-\kappa t}$ versus
  temperature $T$ for different system sizes $L=12,24,48,96$ as in the legend.
  \label{fig:auto_correlation}
}
\end{figure}
In Fig.~\ref{fig:auto_correlation} we show the spin autocorrelation on a $L=96$
system for a range of temperatures $\beta=1,\cdots,10000$ as a function of time
$t$.

In the short time-regime at large temperatures we observe an initial quadratic
regime, followed at very long times by a diffusive tail $A(t) \sim 1/t$ due to
the conservation of the total magnetisation. This is expected at large
temperatures and times \cite{Mueller1988,*Gerling1989,*Mueller1989,Gerling1990},
and has been established for the clean Kagome AFM
\cite{Robert2008,Taillefumier2014}.

For lower temperatures the quadratic regime shrinks (and we do not access
sufficiently large times to see the diffusive tail), and the behaviour crosses
over into a purely relaxational exponential decay $A(t) \sim e^{-\kappa t}$.

We extract the decay rate $\kappa$ from the auto-correlation by fitting an
exponential in the time window $0 < t < 5 \beta$ and for $A(t)>10^{-2}$. The
decay rate $\kappa(T)$ is found to be temperature-dependent as seen in the inset
Fig.~\ref{fig:auto_correlation} showing the decay rate extracted for different
system sizes versus temperature $T$. Whereas the intermediate temperature range
$10<\beta<100$ is consistent with a linear scaling $\kappa(T)\sim T$, at the
lowest temperatures $10^2<\beta<10^4$ the exponent seems to increase to about
$\kappa(T) \sim T^{1.3}$ The upper intermediate linear scaling is consistent
with the behaviour found for the classical spin liquid on the non-disordered
Kagome \cite{Robert2008,Taillefumier2014}, and the pyrochlore lattice
\cite{Conlon2009}, as well as the predictions of the large-N calculations
\cite{Conlon2009,Taillefumier2014}, whereas the $T^{1.3}$ at lowest temperatures
deviates from previously seen behaviour. However, we cannot definitely say that
this defines a new regime, or if there is a further crossover as temperature
approaches $0$.

\subsection{Structure Factor}
Spin correlations are captured by the dynamical structure factor
\begin{equation}
  \mathcal{S}(\vect{q},t) = \left< \vect{s}(\vect{q},t) \cdot \vect{s}(-\vect{q},0) \right>
\end{equation}
where $\vect{s}(\vect{q},t) = \sum_i \vect{s}_i(t) e^{- i\vect{R}_i \cdot
  \vect{q}}$ is the spatial Fourier transform of the spin configuration. Its
frequency transformed version $\mathcal{S}(\vect{q},\omega)$ maps the spectrum
of the dynamical spin-pair correlations, while the quasi-elastic limit
$\mathcal{S}(\vect{q},\omega=0)$ is sensitive to the presence of order in the
system.

\begin{figure}
  \includegraphics[width=.99\columnwidth]{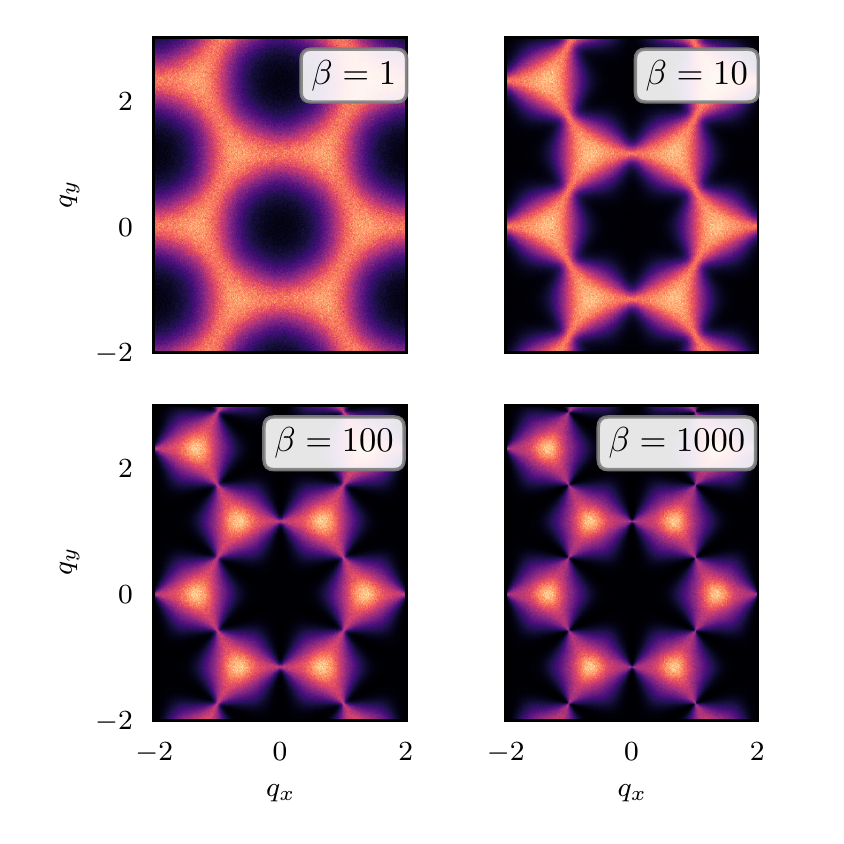}
  \caption{Quasi-elastic structure factor $\mathcal{S}(\vect{q},\omega=0)$ as
    a function of momentum for
    $\beta=1,10,100,1000$ as indicated in the panel.
  \label{fig:static_structure_factor}
}
\end{figure}
\sectionn{Quasi-elastic structure factor} The quasi-elastic structure factor
$\mathcal{S}(\vect{q},\omega=0)$ in momentum space for different temperatures,
$\beta=1,10,100,1000$, is shown in Fig.~\ref{fig:static_structure_factor}. These
temperatures span the regime from paramagnetic down to the fully established
cooperative spin liquid regime for $\beta \gtrsim 10$.

At the largest temperature $\beta=1$ the structure factor only has broad
features in momentum space due to the strong thermal fluctuations in the
paramagnetic state. In the cooperative regime triangular structures of strong
intensity emerge, and with lowering temperature intensity is transferred to the
centers of these regions, which however do not correspond to Bragg peaks as
there is no long-range order. The quasi-elastic structure factor does not change
considerably above $\beta=100$, and does not indicate any long-range order down
to $\beta=10000$, consistent with our previous findings. In particular, note the
absence of the $\sqrt{3}$-satellite-peaks which would be present in the clean
model \cite{Zhitomirsky2008}.

\sectionn{Dynamical structure factor}
\begin{figure}
  \includegraphics[width=.99\columnwidth]{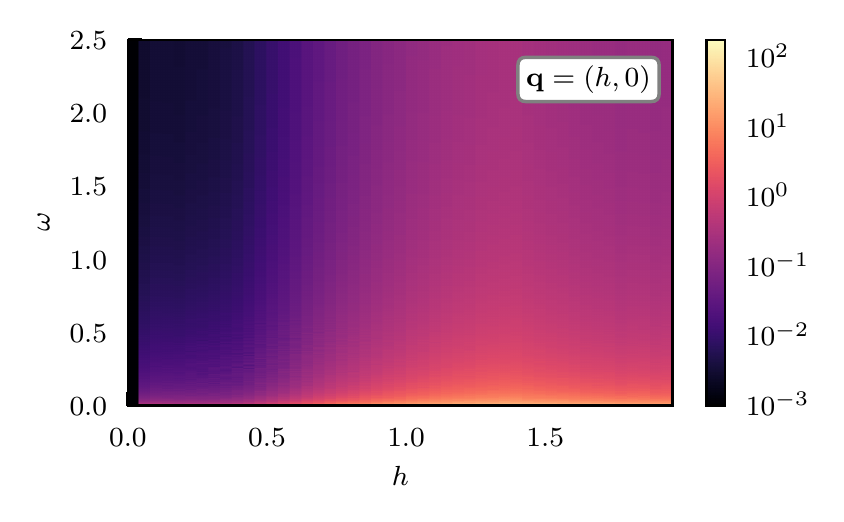}
  \includegraphics[width=.99\columnwidth]{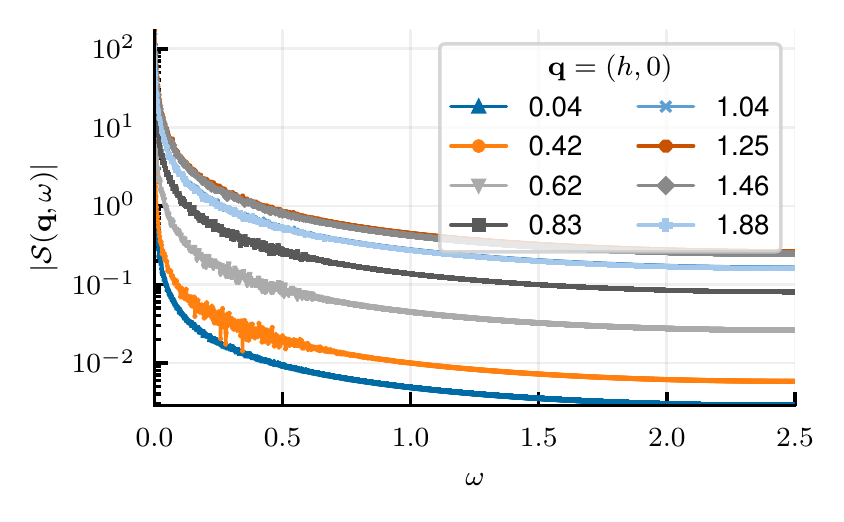}
  \caption{Top: Intensity map of the dynamical structure factor
    $|\mathcal{S}(\vect{q},\omega|)$ versus momentum $\vect{q}=(h,0)$ and frequency
    $\omega$ at a temperature $\beta=100$. Colormap on a logarithmic scale.
    Bottom: Selected cuts of the same data, $\mathcal{S}(\vect{q},\omega)$ as a function of $\omega$ for fixed momenta
    $\vect{q}=(h,0)$ as indicated in the legend.
    \label{fig:structure_factor_w_cut}}
\end{figure}
The dynamical structure factor $\mathcal{S}(\vect{q},\omega)$ only shows broad
features in momentum and frequency space as shown in
Fig.~\ref{fig:structure_factor_w_cut} at a temperature of $\beta=100$ along a
momentum cut from the BZ centre to the edge, $\vect{q}=(h,0)$.

This suggests that there are no sharp spinwaves present in the disordered model,
even at temperatures where they are seen in the clean system \cite{Robert2008}.

In addition, some spectral weight is highly concentrated at small frequencies as
seen in the bottom panel of Fig.~\ref{fig:structure_factor_w_cut}. We associate
this with the large number of soft normal modes discussed below in terms of the
Hessian matrix of ground states.

\sectionn{Diffusion}
\begin{figure}
  \includegraphics[width=.99\columnwidth]{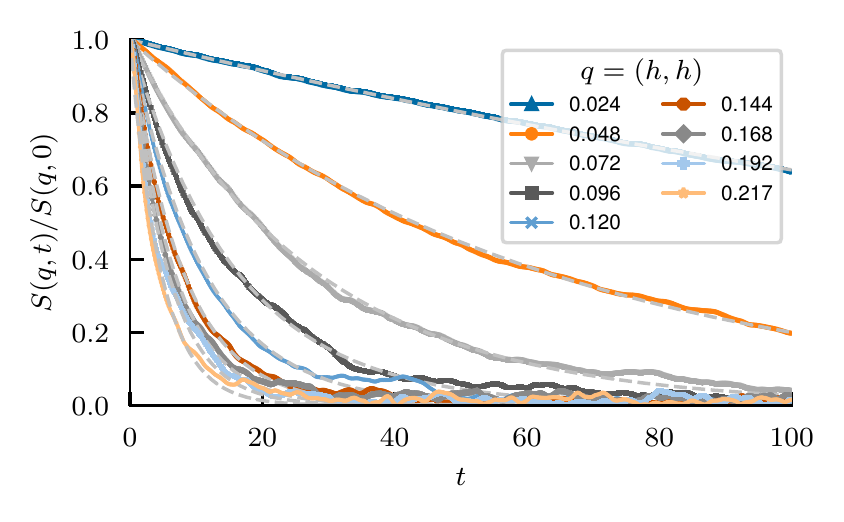}
  \includegraphics[width=.99\columnwidth]{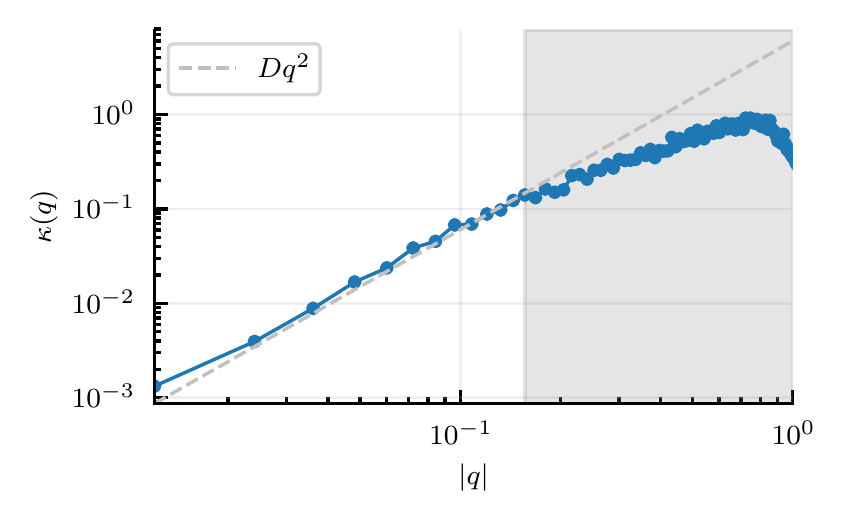}
  \caption{Top: Cuts of dynamical structure factor $\mathcal{S}(\vect{q},t)$ for
    momenta $\vect{q}=(h,h)$ close to the Brillouin zone center versus time $t$ for temperature $T=0.1$,
    dashed gray lines are fits to an exponential decay
    $\mathcal{S}(\vect{q},t) \sim e^{-\kappa(q)t}$
    Bottom: Extracted decay constant $\kappa(q)$ versus momentum $|q|$ with a
    quadratic fit $\kappa(q) = D q^2$ (dashed line), shaded area denotes where quadratic fit
    ceases to be valid.
    \label{fig:structure_factor_cut}}
\end{figure}
Cuts of the dynamical structure factor $\mathcal{S}(\vect{q},t)$ as a function
of time $t$ at fixed momentum $q=(h,h)$ close to the Brillouin zone center are
shown in Fig.~\ref{fig:structure_factor_cut}. We observe an exponential decay in
time $S(\vect{q},t) \sim e^{-\kappa(\vect{q})t}$ with a momentum dependent decay
rate.

The decay rate itself depends quadratically on momentum $\kappa(\vect{q}) = D
q^2$ (lower panel of Fig.~\ref{fig:structure_factor_cut}), at least for
sufficiently small momenta close to the center of the BZ. This in turn allows us
to obtain the diffusion constant $D$.

We note that the range of validity of this quadratic dependence shrinks with
temperature, a behaviour already observed in the clean Kagome magnet
\cite{Taillefumier2014}. In addition, the functional form above this threshold
momentum changes, flattening into a plateau of constant decay-rate. However, we
cannot exclude that diffusion still takes place at smaller wave-vectors, or
larger length scales than we can access in the simulations.

We also note that since the decay rate of the auto-correlation function, which
corresponds to some average of the decay-rates of the momentum-resolved
structure factor, continues to decrease with temperature, the range of the
quadratic behaviour must decrease and/or the the diffusion constant must
decrease at low temperatures.

\begin{figure}
  \includegraphics[width=.99\columnwidth]{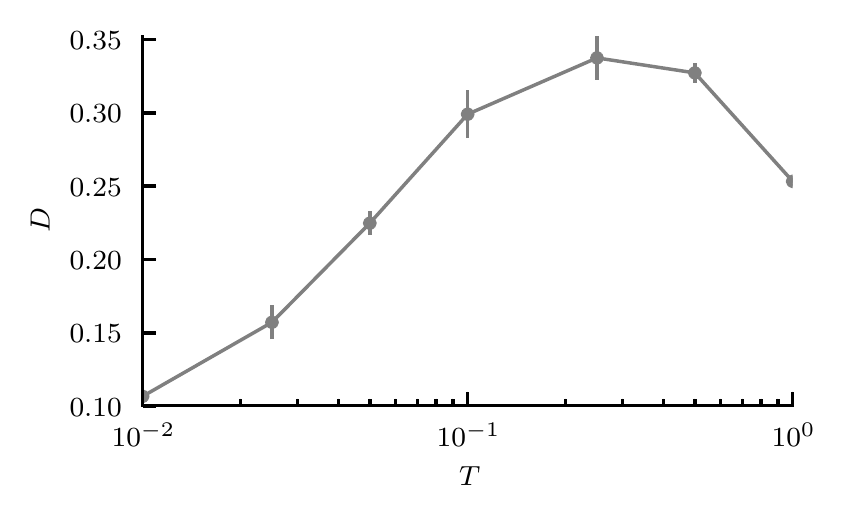}
  \caption{Diffusion constant $D$ as a function of temperature $T$ as extracted
    from the quadratic dependence of the decay-rate of the dynamical structure factor.
    \label{fig:diffusion_constant}}
\end{figure}
The extracted diffusion constant $D$ as a function of temperature $T$ is shown
in Fig.~\ref{fig:diffusion_constant}. Upon lowering the temperature, we first
observe an increase, comparable to the one observed in the clean model on the
transition from high-temperature paramagnetic states to the cooperative
paramagnetic regime.

On further lowering the temperature, the diffusion constant starts to decrease.
This is in stark contrast to the situation of the clean model for which the
diffusion constant after reaching a ``plateau'' in the cooperative regime seems
to diverge on approaching the octupolar regime \cite{Taillefumier2014}. In any
case, the observed decrease it relatively slow, and the data does not allow to
conclude if it will continue down to lower temperatures or if it saturates to a
finite value.

\subsection{Finite-size Transition to Dynamically Arrested States}
On finite systems we observe a transition into a dynamically arrested state. In
the arrested regime the dynamics is stuck close to a single groundstate and does
not explore the full phase space.
Dynamics in this regime can be understood as fast small oscillations around a fixed
state in combination with a slow global precession of all spins.

We characterise the dynamical arrest by considering a modified spin
autocorrelation function $A_{mod}(t)$ obtained by globally rotating all spins of
the time-evolved state such that the first spin $\vect{s}_1(t)$ points in the
same direction as $\vect{s}_1(t=0)$ and the second spin $\vect{s}_2(t)$ lies in
the same plane as $\vect{s}_2(t=0)$.
Intuitively, in this way we remove the zero-energy modes due to the global $SO(3)$
invariance of the Hamiltonian, and the rather trivial dynamics of a rigid
rotation of all spins which should not be considered to lead to a different
state. Since we globally rotate all spins, and then rotate all other spins
around a single spin $S_1$, this leaves the energy invariant.

This was not required for the dynamics discussed above, but becomes so now for
the parameters considered here. At the low
temperatures/energy densities at which we observe the freezing transition the
dynamics has slowed down so much that a global slow
precession of the state masks the internal dynamics of the spin state, whereas
at larger temperatures the internal dynamics are fast enough to be fully
resolved before the global precession becomes relevant.

We either sample states via MC from the Boltzmann distribution at a finite
(small) temperature, or add a (small) energy density to a GS obtained from
numerical minimisation of the energy by rotating all spins slightly in their
local exchange fields.

\begin{figure}
  \includegraphics[width=.99\columnwidth]{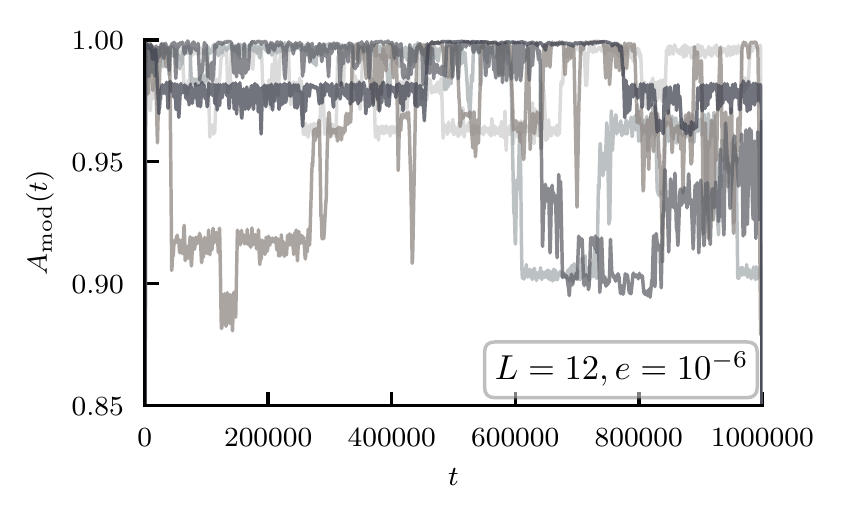}
  \caption{Selected individual time-traces of the modified spin-autocorrelation
    function $A_{\textrm{mod}}(t)$ for a fixed disorder realisation on a $L=12$ system and an added energy
    density $e=10^{-6}$ above different groundstates.
    \label{fig:A_long_time_traces}
  }
\end{figure}
For illustrational purposes we begin by discussing individual time-traces at a
fixed disorder realisation of the
modified autocorrelation function at low energy densities above a ground state
close to the dynamic arrest in Fig.~\ref{fig:A_long_time_traces}. Note in
particular the extremely large times up to $t=10^{6}$ over which we resolve the
dynamics here. We emphasise that these are fixed disorder
trajectories at smaller energy densities and on a different time-scale than the disorder-averaged
spin-spin autocorrelation results in Fig.~\ref{fig:auto_correlation}, which still would
have fully decayed by these times if the previously observed scaling did persist
down to these energies.

In stark contrast to the previously discussed (exponentially) decaying
autocorrelation, here we observe rapid oscillations around a fixed plateaus for long time periods separated by
rapid and sudden transitions to different plateaus.
We interpret this behaviour as the system  system being stuck close to distinct ground
states as characterised by the distinct plateaus for long times, around which it
performs small normal mode oscillations, until a sudden and sharp transition to different plateau/state occurs.

Furthermore, in this regime the system remains close to the original state, in
that we observe transitions out of and back into the original state, and in some
cases repeatedly to the same distinct plateau/state. This would be exceedingly
unlikely if the dynamics were to explore the full exponentially large
groundstate manifold of the jammed spin liquid.

This behaviour is somewhat reminiscent of (finite) spin glass systems which are
stuck for (exponentially) long times in some part of phase space, but may
suddenly jump to a distinct region \cite{Parisi2006}. These distinct states
might also lend themselves to the interpretation of two-level systems,
as observed in Heisenberg spin glasses \cite{Baity-Jesi2015}, and do indicate
some form of clustering of the ground states.

\begin{figure}
  \includegraphics[width=.99\columnwidth]{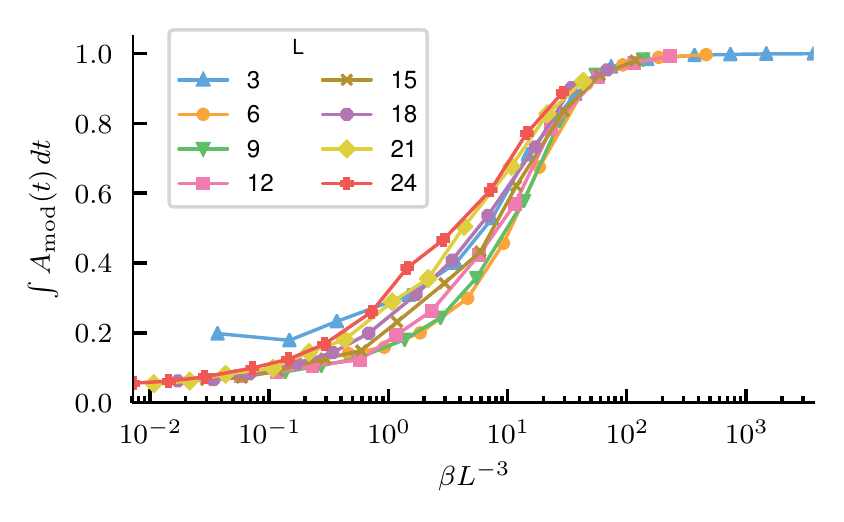}
  \includegraphics[width=.99\columnwidth]{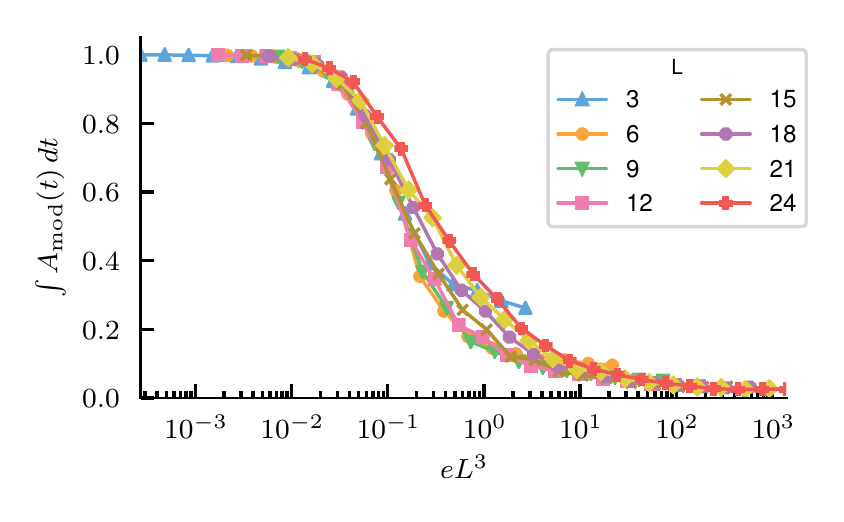}
  \caption{Top: Long time average of the (modified) spin autocorrelation
    function $A_{mod}$ 
    versus scaled temperature $\beta L^{-3}$ in the MC simulations.
  Bottom:Long time average of the (modified) spin autocorrelation
    function of groundstates with an added energy density $e$ created by randomly
    rotating spins in their local exchange fields.
    \label{fig:A_long_time}
  }
\end{figure}
The long-time average of the modified autocorrelation function averaged over
disorder realisations is shown in
Fig.~\ref{fig:A_long_time}, averaged over initial states obtained from
MC-simulations at a finite temperature (top panel), and as a function of energy
density added to a true groundstate ($E=0$) by randomly rotating the spins in
their local exchange field (bottom panel).
 (We have checked that the same transition with the same scaling occurs for the
 random bond model.)

The finite temperature Monte Carlo results display a clear crossover as a
function of temperature between dynamics which explores (some of) phase space
and $A_{mod} \approx 0$, and dynamics at low temperatures which is stuck near a
single groundstate with $A_{mod} \approx 1$. Similarly, the groundstate
simulations show a transition as a function of added energy density with the
same scaling.

We note that the transition to a dynamically arrested state appears to be a
finite size effect, in that the temperature below which the dynamics is arrested
scales as $T \sim L^{-3}$ for the MC simulations, and energy density $e \sim
L^{-3}$ for the groundstate simulations. This leads us to conclude that the
energy barriers between different JSL groundstates vanish in the thermodynamic
limit.

\section{Hessian\label{sec:hessian}}
To elucidate the behaviour found above, we first investigate the statistical
properties of an individual local extremum, before turning to their connectivity
properties in the following sections.

Following on from our original work \cite{Bilitewski2017}, we investigate the
quadratic energy cost of fluctuations around groundstate configurations via the
Hessian matrix. This provides insight into the spectrum of fluctuations,
potential low-energy or zero-modes, and via the associated eigenvectors also
into the spatial properties of these normal modes.

For a spin configuration $\left\{\vect{s}_i \right\}$ we choose an orthonormal
local basis at every lattice site $(\vect{s}_i, \vect{u}_i, \vect{v}_i)$. This
allows us to parametrise fluctuations as $\vect{\tilde{s}}_i =
\sqrt{1-\epsilon_i^2} \vect{s}_i+ \epsilon_{ui} \vect{u}_i + \epsilon_{vi}
\vect{v}_i$ with $\epsilon_i =(\epsilon_{ui},\epsilon_{vi})$ which takes the
spin normalisation condition into account. Around a groundstate the energy-cost
of fluctuations to quadratic order is then given by $E = \epsilon^T M \epsilon
$, which defines the $(2N_s) \times (2N_s)$ Hessian matrix $M$.

\begin{figure}
  \includegraphics[width=.99\columnwidth]{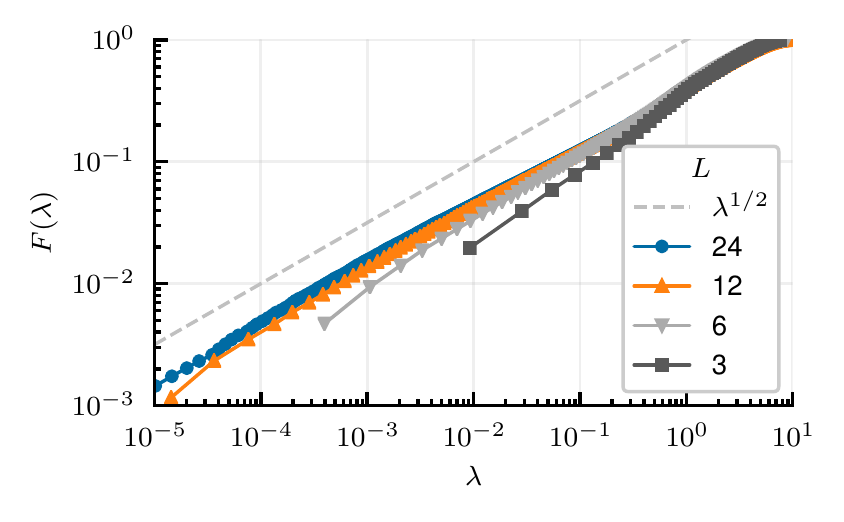}
  \includegraphics[width=.99\columnwidth]{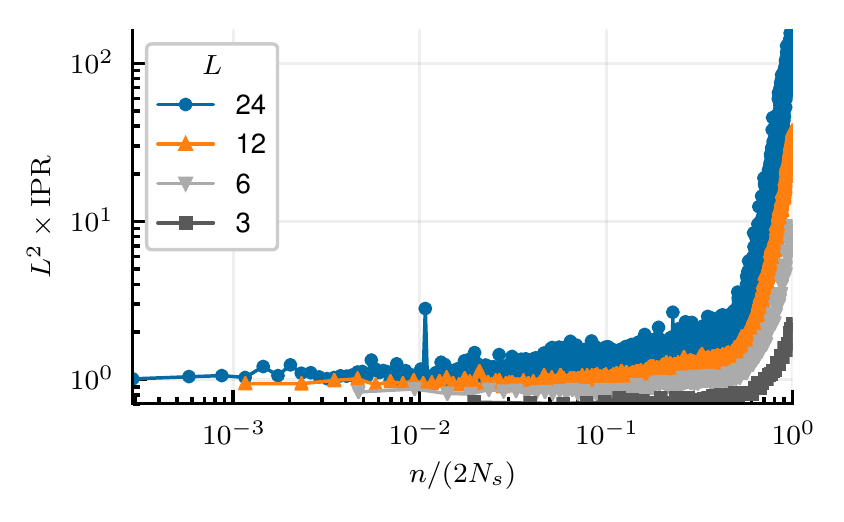}
  \includegraphics[width=.99\columnwidth]{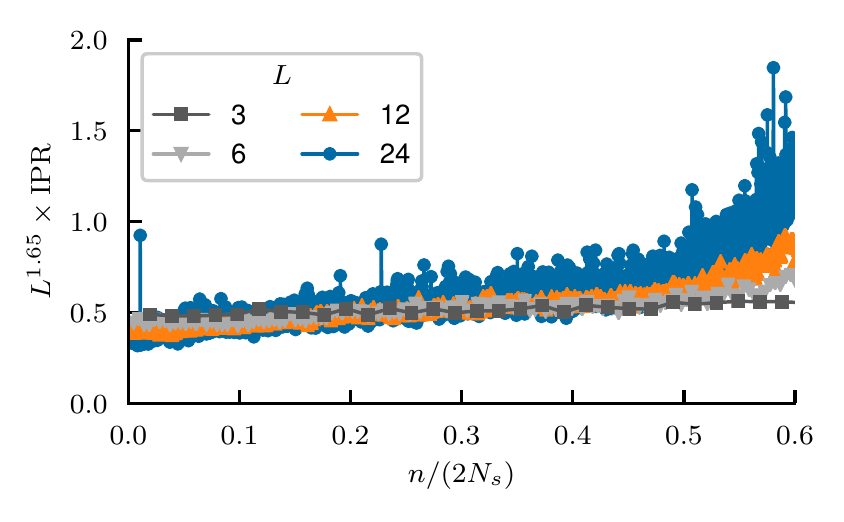}
  \caption{Top: Cumulative distribution function $F(\lambda)$ of the
    eigenvalues $\lambda$ of the Hessian matrix of groundstates for different
    linear system sizes $L$ as indicated in the legend, and $\lambda^{1/2}$
    powerlaw behaviour (dashed) line
    Middle and Bottom: Inverse participation ratio ($\textrm{IPR}$) of eigenstate $n$ of the
    Hessian ordered by eigenvalue, middle scaled with $L^2$ on log-log scale,
    bottom scaled with $L^{1.65}$ on linear scale focussing on the soft modes.
    \label{fig:hessian}}
\end{figure}

Diagonalising the Hessian matrix $M$ provides eigenvalues $\lambda$ and the
corresponding eigenmodes. Due to the global rotational invariance of the energy
there are 3 trivial zero-modes which we do not consider below.

We analyse the spectrum by considering the cumulative distribution function
$F(\lambda)=\int_0^{\lambda} P(x) \,\textrm{d}x$ of the Hessian eigenvalues
averaged over disorder which is shown in Fig.~\ref{fig:hessian}. Note that this
has the advantage of being mostly independent of system size, with larger
systems simply extending the results down to smaller eigenvalues. We observe a
large number of soft-modes with a low energy scaling $F(\lambda) \sim
\lambda^{1/2}$. In that sense the jammed spin liquid states are marginally
stable, as soft modes extend as a powerlaw to zero energy. Crucially, there are
no non-trivial zero-modes, in contrast to the coplanar states of the clean
Kagome system which hosts an extensive number of these.

Secondly, we study the localisation properties of these modes by considering the
inverse participation ratio (IPR) \cite{Mazzacurati1996, Bell1970} defined as
\begin{equation}
  \textrm{IPR} = \sum_i{\left(\left| \epsilon_i \right|^2\right)^2}/\sum{\left|
      \epsilon_i \right|^2}\, ,
  \label{eq:IPR}
\end{equation}
which is 1 for an eigenmode fully localised on a single site of the lattice, and
$1/N_s\sim L^{-2}$ if it is fully delocalised.

The IPR is shown in the bottom panels of Fig.~\ref{fig:hessian} for different
linear system sizes $L$. We observe a tendency towards delocalisation for most
of the spectrum, in particular for the soft modes, with a best fit fractal
exponent $\textrm{IPR} \sim L^{-5/3}$. In contrast, the ``hardest'' modes at the
upper edge of the spectrum are strongly localised to a few sites.

This is notable since the coplanar states of the clean model have an extensive
number of localised zero-energy modes, in particular the
$\sqrt{3}\times\sqrt{3}$ state admits hexagon weatherwane modes involving only 6
sites, and the $q=0$ state admits modes which involve $L$ sites.

These results confirm the picture that the ground states of the jammed spin
liquid have no non-trivial zero-modes, but a large number of relatively soft
modes. Interestingly, these soft-modes appear to be delocalised over the full
lattice, rather than being local excitations like in the clean model.

\section{Forcing / Spectroscopy of Energy Barriers\label{sec:forcing}}
The results on the dynamics indicated that at sufficient energy/temperatures the
system can explore a large part of phase space, whereas at low energies finite energy barriers
between distinct ground states, which scale to zero in the infinite system size
limit, inhibit dynamics freezing the system close to one ground state.
Furthermore, the study of the Hessian showed that each local minimum has no
non-trivial zero-modes, thus, locally appearing as a quadratic well in
configuration space. We now set out to explicitly probe the energy barriers
between distinct close groundstate configurations.

\subsection{Method\label{sec:forcing_method}}
To explore the groundstate manifold further and gain insight into the
energy-barriers between distinct groundstates we use the following protocol,
which we adapt from its application in the study of spin glasses
\cite{Baity-Jesi2015}.

\begin{itemize}
\item Find a groundstate $\mathrm{GS}$ of original hamiltonian $\mathcal{H}$ (Eq.~\ref{eq:H_general})
\item Find a groundstate/local minimum $\mathrm{GS}(h) $ of a perturbed hamiltonian
  $\mathcal{H}_{h}$ (defined in Eq.~\ref{eq:H_pert}) with
  a force/magnetic field $h$ added starting from $\mathrm{GS}(0)$
\item Find a groundstate $\mathrm{GS}^{*} $ of original hamiltonian
  $\mathcal{H}$ starting from $\mathrm{GS}(h)$
\item Compare the newly obtained groundstate $\mathrm{GS}^{*}$ with the original
  state $\mathrm{GS}$
\end{itemize}
\begin{figure}
  \includegraphics[width=.99\columnwidth]{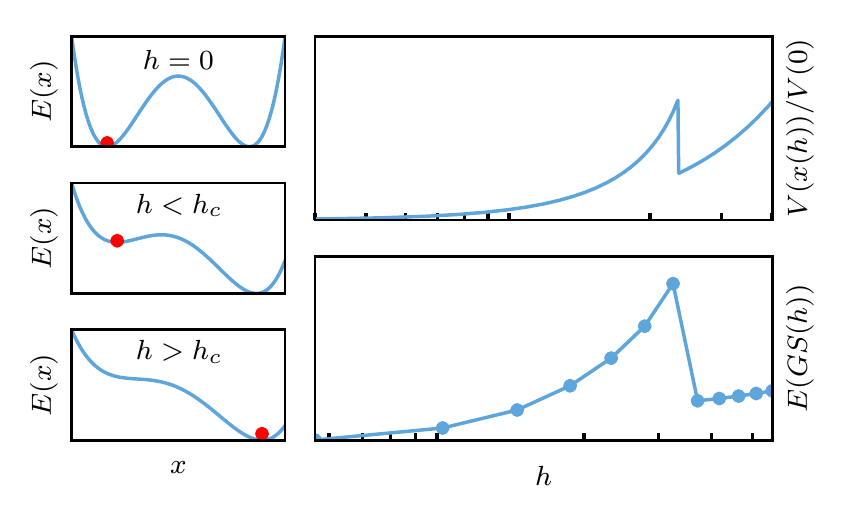}
  \caption{Illustration of the proposed forcing protocol (see text). Left panels:
    one-dimensional double well potential $V(x)$ with an added linear force
    $E(x)=V(x)+ h F(x) $ proportional to $h$. Red circle denotes the state $x(h)$ at the minimal
    energy of the perturbed potential. Right panels: energy versus forcing strength; top for the
    one-dimensional toy model, bottom energy $E$ computed from $\mathcal{H}$ of
    the perturbed state $\mathrm{GS}(h)$ for a single realisation of the proposed forcing protocol in the full spin system.
    \label{fig:forcing_toy}}
\end{figure}

Intuitively, this protocol can be motivated using a one-dimensional analogy as
illustrated in Fig.~\ref{fig:forcing_toy}. Imagine a
particle in a double well potential (corresponding to two distinct ground states
of the jammed spin liquid which act as quadratically confining wells). Starting
with the particle in one well, one can add a linear force to make it move up the
barrier towards the other well (corresponding to state $\mathrm{GS}(h)$).
As long as the particle does not cross the maximal height of the barrier between
the wells after removing the linear force it will fall back
to the initial state (e.g. $\mathrm{GS}^{*}=\mathrm{GS})$ as seen in the middle
left panel of Fig.~\ref{fig:forcing_toy}. At sufficently large applied force the
particle reaches the maximum of the potential between the two minima, at larger
applied force it will then fall into the next well (e.g. $\textrm{GS}* \neq
\textrm{GS}$ ) as seen in the lowest left panel of Fig.~\ref{fig:forcing_toy}.
The minimal value of the force required for this to happen then
defines $h_c$ and the height of the barrier corresponds to the energy of the
particle at $h_c$.

The right panels of Fig.~\ref{fig:forcing_toy} compare the resulting energy
observed at a certain strength of the forcing for the toy model (top) to one
realisation of the forcing protocol for the full spin-model (bottom) demonstrating qualitative agreement
with this analogous model.

We define the Hamiltonian $\mathcal{H}(h)$ as
\begin{equation}
  \mathcal{H}(h) = \mathcal{H} + h \sum_i \vect{s}_i \cdot \vect{h}_i
  \label{eq:H_pert}
\end{equation}
where we choose the magnetic fields $\vect{h}_i$ to be orthogonal to the initial
groundstate $\mathrm{GS}$, i.e. $\vect{h}_i \cdot \vect{s}^{\mathrm{GS}}_i=0$,
and normalised as $\sum_i \vect{h}_i^2 =1$. (This is to say that the
$\{\vect{h}_i\}$ form a normalised element in the tangent space of
$\mathcal{S}_2^{N_s}$)

By choosing the field local and in the tangent space we avoid the issue that due
to rotational invariance of the hamiltonian the main response of any state to a
global field will just be to align with the field direction.

We consider different scenarios for this forcing: (a) we choose the direction of
the forcing to correspond to the softest direction of the Hessian matrix of the
initial groundstate $\mathrm{GS}$, (b) the hardest direction of the Hessian, and
(c) a random direction in the tangent space of the initial groundstate
$\mathrm{GS}$.

We emphasise that this protocol inherently goes beyond the linear response
regime which would be fully captured by the eigenvalues of the Hessian. The
purpose is to perturb the state strongly enough to leave the local basin of
attraction of the initial state resulting in a (potentially sudden) non-linear
response.

In addition, it allows us to extract (local) information about the set of
groundstates which is not accessible from the states alone. Namely, we will
obtain the critical fields and the height of energy-barriers between
``neighbouring'' (those connected by the protocol above) groundstates, and the
locality of changes between these ``neighbouring'' states.

We note that the perturbed state $\mathrm{GS}(h)$ we obtain numerically is not necessarily a ground
state of $\mathcal{H}(h)$, but rather only a local minimum. However, we are
actually not interested in the ground states of $\mathcal{H}(h)$ in any case, as
we only use it to perturb the original ground states in a deterministic fashion. Further starting from $\mathrm{GS}(h)$ we are not guaranteed to obtain a
true groundstate $\mathrm{GS}^{*}$ of $\mathcal{H}$ with $E=0$, but may also end up in a local
minimum. These cases are however easily distinguished by the non-vanishing of
the energy and for the discussion below we only consider cases for which the
minimisation results in a global minimum.


\subsection{Critical forcing strength}
To find the critical field $h_c$ required to leave the basin of attraction of a
given groundstate $\mathrm{GS}$, we follow the protocol outlined above. Thus, we
initialise $h$ at a very small value, compute the perturbed state $\mathrm{GS}(h)$ and associated state
$\mathrm{GS}*$, and increase $h$ until we encounter a new state $\mathrm{GS}^{*} \neq
\mathrm{GS}$ for the first time, as characterised by an overlap with the inital
state unequal to 1.

In practice starting from a ground state obtained via energy minimisation at $h=0$, we
start with a small field $h \sim 10^{-7}$, increase it in powers of 10 until we
find a different state, and then perform a refined search between the last two
values of $h$ to determine $h_c$. Deciding whether a new state is encountered
during this procedure poses no numerical problems as using the overlap
$q$ proves sufficient given the convergence criteria put on the states (though the authors have also checked the results comparing the full gram
matrix $g_{ij} = \mathbf{s}_i \cdot \mathbf{s}_j$ which is in one-to-one
correspondence to spin configurations modulo global rotations).

\begin{figure}
  \includegraphics[width=.99\columnwidth]{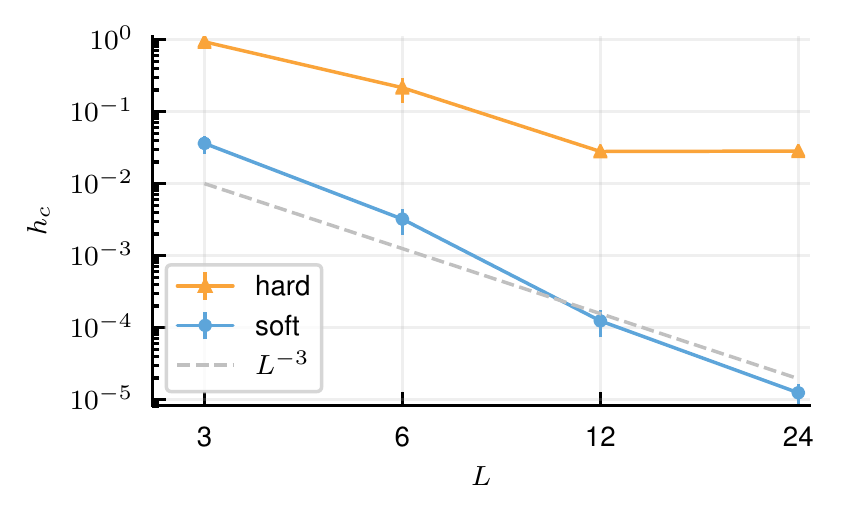}
  \caption{Comparison of critical force required to leave a groundstate for
    forcing in the softest direction (blue circles)
    and the hardest direction (orange triangles). Dashed line is a guide to the
    eye powerlaw $\sim L^{-3}$.
    \label{fig:critical_force}
  }
\end{figure}

The results for forcing in the softest and the hardest direction are shown in
Fig.~\ref{fig:critical_force} with errors obtained from the average over 100 different disorder
realisations/initial states as a function of linear system size $L$. For
forcing in the softest direction we observe a vanishing of the critical field
strength with system size consistent with a $L^{-3}$ scaling (dashed line).

For forcing in the hard direction the critical force first decreases, but then
saturates for system sizes $L\ge 12$ at a finite value. Note also the order of
magnitude difference between the critical fields.

\subsection{Energy Barriers}
In addition to the critical force required to leave a GS we may estimate the
energy barrier between the different GS in the follwing way: As we increase $h$
we obtain a series of states $\mathrm{GS}(h)$ and associated $\mathrm{GS}^{*}$,
at some critical $h_c$ the new state $\mathrm{GS}^{*}$ differs from the initial
groundstate. We estimate the energy barrier between $\mathrm{GS}$ and
$\mathrm{GS}^{*}$ by the bond-energy of the state $\mathrm{GS}(h)$, i.e. its
energy with respect to $\mathcal{H}_0$, for $h$ just below the critical field
$h_c$.

\begin{figure}
  \includegraphics[width=.99\columnwidth]{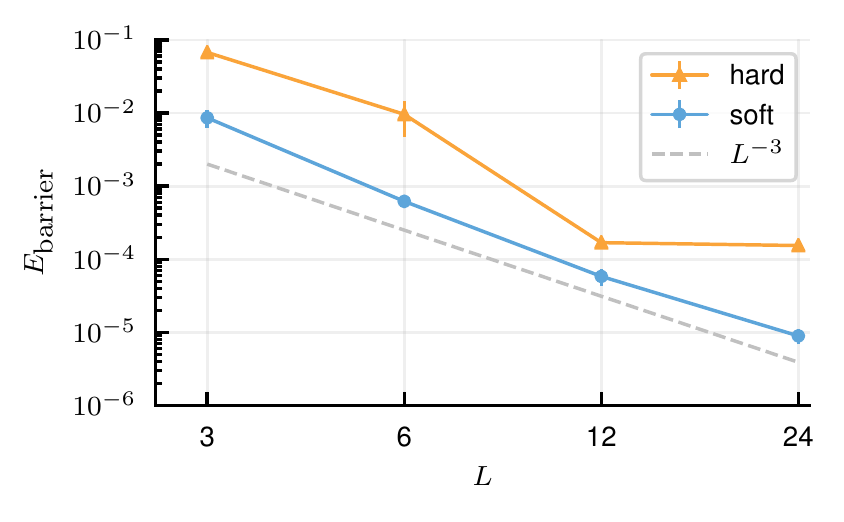}
  \caption{Comparison of energy-barriers between groundstate
    configurations for forcing in the softest direction (blue circles)
    and the hardest direction (orange triangles). Dashed line is a guide to the
    eye powerlaw $\sim L^{-3}$.
    \label{fig:energy_barrier}
  }
\end{figure}

The results of this are shown in Fig.~\ref{fig:energy_barrier} for forcing in
the softest direction and forcing in the hardest direction as a function of
system size/number of spins.

We observe that for forcing in the softest direction the height of energy
barriers decreases with system size as $L^{-3}$, implying that transitions
between states are possible at thermodynamically vanishing energy cost.

In contrast, forcing in the hardest direction states faces a finite energy
barrier which appears to saturate on larger systems consistent with the observed
behaviour of the critical fields.

Whereas we cannot fully exclude any potential bias stemming from the initial
ground states at larger systems which are harder to converge numerically, both the fact that the
statistical errors actually decrease with increasing system size and the
consistency of the ground state with Monte-Carlo simulations combined with the
orders of magnitude difference between ``soft'' and ``hard'' forcing leads us to believe
that this is a robust effect.

Based on the results for the critical force and the associated energy barriers,
it appears that while states on finite systems have no zero-energy modes,
transitions can be induced by vanishingly small forces and at vanishingly small
energy cost if the force is applied in the right direction, in keeping with our
results above on the stability of the arrested regime.

\subsection{Response of states to forcing}
We can also characterise the response of the state to the introduced forcing by
measuring its magnetisation along the applied magnetic field
\begin{equation}
  m =\braket{\mathrm{GS}(h)}{h} =\sum_i \vect{s}_i(h) \cdot \vect{h}_i
\end{equation}

\begin{figure}
  \begin{minipage}[t]{0.99\columnwidth}
    \vspace{0pt} \includegraphics[width=.99\columnwidth]{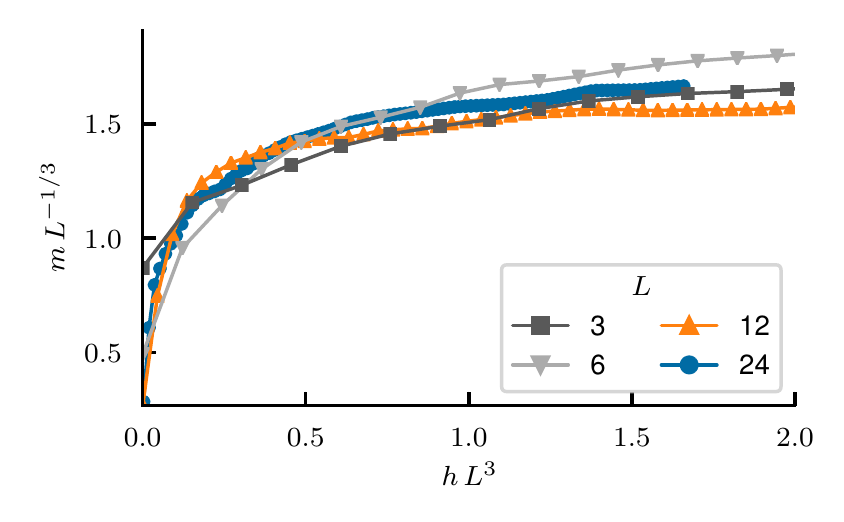}
    \includegraphics[width=.99\columnwidth]{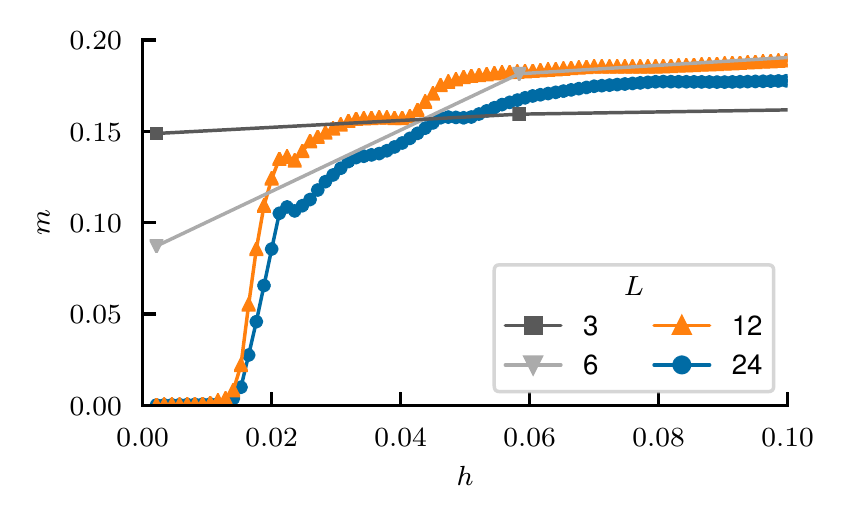}
  \end{minipage}
  \caption{Comparison of the response to the applied force, i.e. the induced
    magnetisation, of a groundstate versus forcing strength for forcing in the
    softest direction (top panel)
    and the hardest direction (bottom panel). Note the different ranges and the
    differently scaled axes.
    \label{fig:response}
  }
\end{figure}

In Fig.~\ref{fig:response} we again compare the forcing in the softest direction
with the forcing in the hardest direction for different system sizes $L$. Note
that as per the observed scaling of the critical fields above we scale the
magnetic field with $L^3$ for forcing in the soft direction, and the resulting
response by $L^{-1/3}$ to collapse data for different system sizes.

For forcing in the soft direction we observe a continuous response to the
applied field. Because the energy landscape is extremely shallow in direction of
the smallest eigenvector of the Hessian, the state shows a strong response to
the applied field as it smoothly moves along the bottom of local basin of
attraction of the initial state.

In contrast for forcing in the hard direction we observe two qualitatively
distinct regimes: weak response at small fields, and above a crossover field a
rapid increase of the induced magnetisation. In addition, the response is
smaller in magnitude than for the soft forcing direction as expected as now the
state moves along a steep direction in energy.

The observation of a ``gapped'' response for forcing in the hard direction is
consistent with the finite critical forces and energy barriers observed above.
If the field is too small to leave the initial basin of attraction, responses
are weak along the steeply confined direction in configuration space, whereas
when exceeding a critical field the perturbed state can escape the
initial state, showing an abrupt response. After this sudden response the spin
configuration ends up in a distinct state for which the forcing direction might
not corespond to a strongly confined direction anymore, and for which the field
direction is also not perpendicular to the state anymore, thus completely
changing it's response.

We conclude that ground states show a strongly anisotropic behaviour with order
of magnitude differences in the response depending on the direction of the
applied force.

\subsection{Overlaps}
We next turn to characterise the ``neighbouring'' states in more detail,
beginning with overlaps or distances between these states, allowing us to draw
conclusions on the clustering of groundstates.

We compare the original groundstate $\mathrm{GS}$ and the first different
groundstate $\mathrm{GS}^{*}$ encountered when increasing the forcing strength
via their average overlap $q$,
\begin{equation}
  q = \frac{1}{N_s} \sum_i \vect{s}_i \cdot \vect{s}_i^{*}\, .
\end{equation}
Again, we define this after rotating both states into a standard form with
$\vect{s}_1=\vect{e}_z$, and $\vect{s}_2$ in the $xz$-plane exploiting the
rotational invariance.

This provides a global notion of the total change required to transition from
one GS to another, and since $|\vect{s}_i - \vect{s}_i^{*}|^2 = 2 - 2\, \vect{s}_i \cdot \vect{s}_i^{*}$
 also geometrically corresponds to the distance between states, providing
 complementary information to the physical energy barriers and critical forcing fields discussed above.

\begin{figure}
  \includegraphics[width=.99\columnwidth]{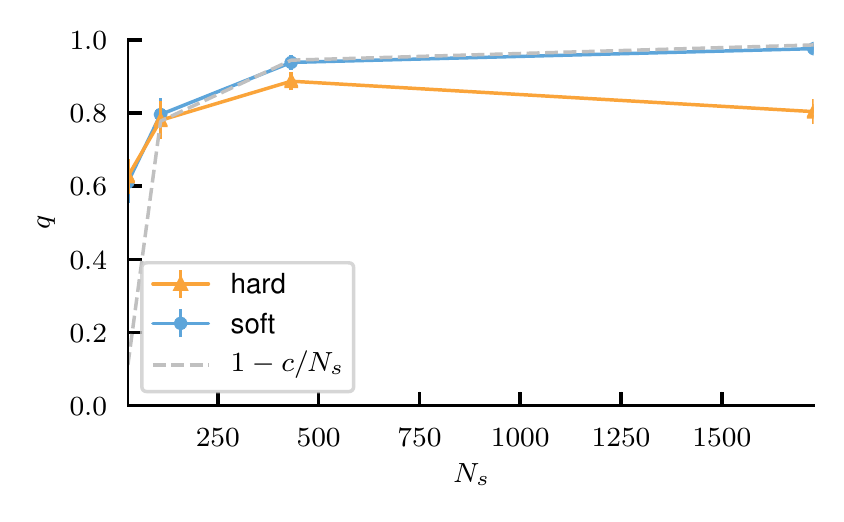}
  \caption{Comparison of overlaps for forcing in the softest direction (blue circles)
    and the hardest direction (orange triangles). Average overlap per spin $q = 1/N_s
    \sum_i \vect{s}_i \cdot \vect{s}_i^{*}$.
    \label{fig:overlaps}
  }
\end{figure}
In Fig.~\ref{fig:overlaps} we observe that forcing in a soft direction leads to
a state $\mathrm{GS}^{*} $ with a high overlap with the original state
$\mathrm{GS}$. This increases with increasing system size, converging towards
$1$ as $q=1-c/N_s$ with a constant $c \approx 24$.
This seems to suggest that, on average, only a constant number of rearrangements is required to transition
into a ``neighbouring'' ground state, but as discussed below these
rearrangements are in fact not localised, but rather require a change of all spins in
the spin configuration. Thus, this result only indicates the there exist
many groundstates close by.

In contrast, forcing in the hardest direction results in an overlap $q<1$, which
actually tends to decrease on larger systems. However, we note that this is
still relatively large considering that there are exponentially many
groundstates of the JSL, and for a random new state we would expect $q \approx
0$.

This fits the interpretation that for forcing in the soft direction the GS
smoothly evolves moving along a shallow basin of the energy landscape with an
associated gradual change of the spins, whereas for forcing in a hard direction
the evolution is along a steep direction with a sudden transition into a new
basin of attraction resulting in a more strongly perturbed final spin configuration.

\subsection{Localisation of changes}
Finally, we consider the locality of rearrangements required to change one
groundstate into the other.

For the coplanar states of the clean model, in particular the $\sqrt{3}\times
\sqrt{3}$ state, there are local zero-energy normal-modes that allow to move
within the ground state manifold. For the non-coplanar groundstates of the
disordered model this is not the case any longer. However, we have observed
above that for forcing in a soft direction only a small change in the
spin-configuration is required. Thus, it is natural to ask how this change is
distributed over the lattice.

We define as the measure of localisation
\begin{equation}
  w_{loc} = \sum_i{w_i^2}/\left(\sum{w_i}\right)^2
\end{equation}
with $w_i = 1-\vect{s}_i \cdot \vect{s}_i^{*}=(\vect{s}_i-\vect{s}_i^{*})^2/2$,
analogous to the IPR discussed for the normal modes of the Hessian. It is $1$ if
the change is fully local and only a single spin is changed, and $1/N_s$ if the
change is homogeneously delocalised over all $N_s$ spins.

\begin{figure}
  \includegraphics[width=.99\columnwidth]{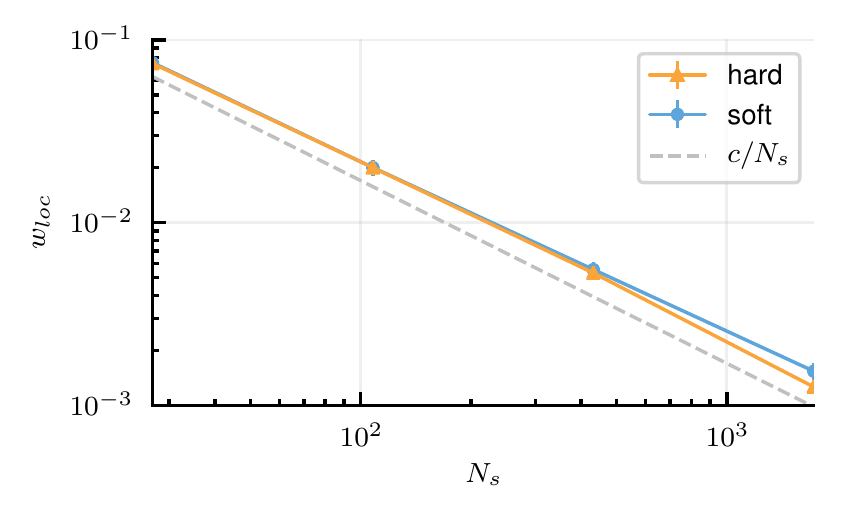}
  \caption{Comparison of localisation of changes between groundstate
    configurations for forcing in the softest direction (blue circles)
    and a random direction (orange triangles).
    \label{fig:localisation}
  }
\end{figure}
We show the results for forcing in the soft and the hard direction in
Fig.~\ref{fig:localisation}. In both cases we observe a scaling $w_{loc}\sim
1/N_s$ corresponding to changes of the spin configuration delocalised over the
full lattice.

We note that this is in agreement with the nature of the soft modes of the
Hessian which we also found to be delocalised over the full lattice. However, it
is in contrast to the behaviour observed in Heisenberg spin glasses, where local
rearrangements between different low-lying states exist and have been found
using a similiar protocol \cite{Baity-Jesi2015}.

To reconcile the fact that the overlap suggests on average only a small number of
changes in the spin-configuration and the delocalisation of this change over the
full lattice recall that the non-disordered model has localised zero-modes and
that the groundstates are defined via the set of strict constraints on each triangle of the lattice. 
If one now changes one spin locally, one has to change all spins in the
triangles it belongs to to compensate. In turn all spins in the neighbouring
triangles have to be adjusted to satisfy the constraint, continuing throughout
the full lattice.
It is rather remarkable that in the clean model this series of changes
terminates and can be localised, whereas for the disordered spin groundstates in
presence of disordered constraints it appears to require a global, but small
change in the spin configuration.

\subsection{Local Forcing}
Finally, we also consider a local perturbation to see whether the non-locality
of the re-arrangements observed above might have been due to our globally
applied field, rather than an inherent property of the probed states.

Specifically, we choose $\vect{h}_i \neq 0$ only on a single triangle $\alpha$
with a field direction $\vect{h}_i$ choosen randomly on the unit sphere. Note
that applying a field local to a single spin only would, due to the rotational
invariance of the field-free Hamiltonian, just lead to a global rotation of the
state into field-direction,
such that at least two fields are required to induce a non-trivial response.
Further, a single applied field still leaves the zero mode of rotation around
that field, thus even applying two local fields one encounters a zero mode.
Thus, we choose a single triangle with three spins as the smallest local unit
which avoids these issues.

\begin{figure}
  \includegraphics[width=.99\columnwidth]{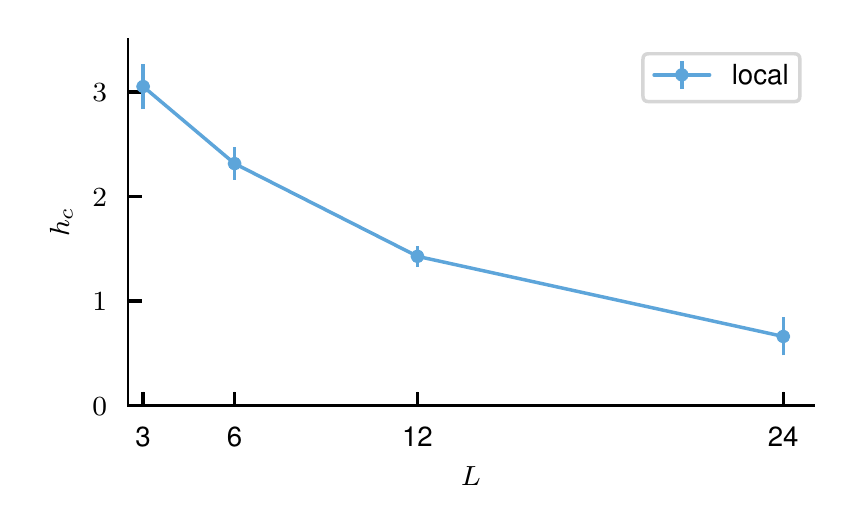}
  \includegraphics[width=.99\columnwidth]{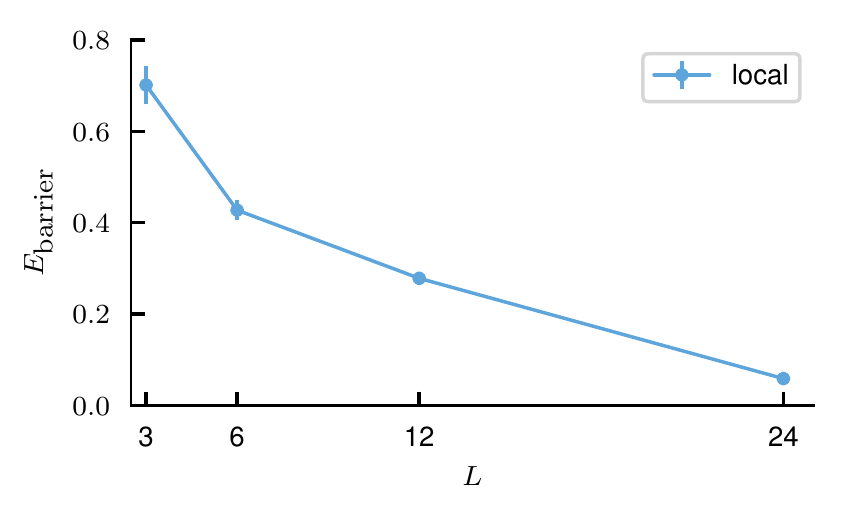}
  \includegraphics[width=.99\columnwidth]{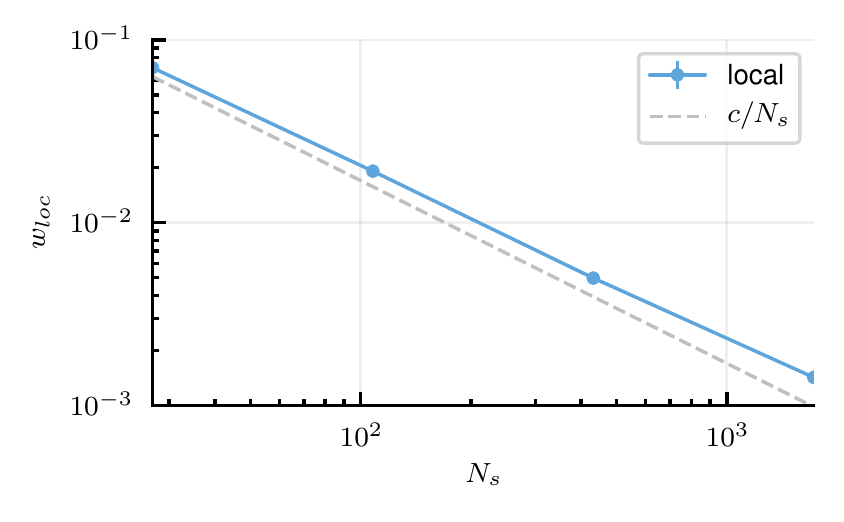}
  \caption{Top: Critical force required to leave a groundstate when applying a
    local field $h$ to a single triangle oriented in a random direction, middle: Corresponding energy barrier, both
    versus linear system size $L$ bottom:
    $w_{loc}$ of rearrangements between ``neighbouring'' groundstates, versus
    number of spins  $N_s$
    \label{fig:local_critical}
  }
\end{figure}
We again first consider the critical field strength and energy barriers in
Fig.~\ref{fig:local_critical}. We observe that even though they decrease with
increasing system size, they are substantially larger than for the globally
applied perturbations. Whereas we cannot make a precise statement on the
asymptotic value, the fact that the Hessian of the groundstates did not have
localised soft modes, and the constraints within the groundstate manifold appear
locally rigid, strongly suggests that this energy cost remains finite.

The induced rearrangements of the spin-configurations after exceeding the
critical field again appear to be de-localised over the full lattice (bottom
panel of Fig.~\ref{fig:local_critical}), in spite of the local nature of the
applied field, lending additional confidence to the conclusion that such
non-local changes are in fact required to transition to a distinct ground state.

We note that this is different from the coplanar states of the clean model,
which are unstable to an infinitesimal out-of-plane local perturbation due to
the local zero-modes. Indeed performing the same protocol on coplanar states of
the clean model with local zero-modes the critical field as well as the energy
cost vanishes, in addition to observing a finite localised response for infinitesimal applied field
(as long as the out-of-plane component of the applied field is non-zero).

Thus, at least within the protocol described we do not find any local soft modes
of the jammed spin liquid that would allow transitions between different ground
states.

\section{Random Walk in groundstate space\label{sec:random_walk}}
The discussion above provided information on the local properties of the set of
groundstates, critical fields and energy barriers between ``neighbouring''
states. Next, we consider potential clustering and the size of the basins of
attraction of different ground states, and the ``connectedness'' of the ground
states.

To this end we propose starting from a given initial GS $\{S_i\}_0$ to
repeatedly apply the procedure above to generate a sequence of states
$\{S_i\}_n$, which may either be local minima or true GS, e.g.
\begin{itemize}
  \item Find an initial ground state $\mathrm{GS}_0$ of $\mathcal{H}$
  \item in step $i$: Apply a perturbing field $h$ as detailed in
    sec.~\ref{sec:forcing_method} starting from $\mathrm{GS}_{i-1}$, increasing the field strength until
    $\mathrm{GS}^{*}_i\neq \mathrm{GS}_{i-1}$, obtaining a new distinct state $\mathrm{GS}_i=\mathrm{GS}^{*}_i$
  \item repeat this $n$-times obtaining a sequence of states
\end{itemize}

In this case we mainly discuss applying the force in a random direction. Forcing
the states exclusively in the softest or hardest direction as discussed above,
the procedure can get stuck in a trivial cycle, typically consisting of two
states connected by a steep/shallow direction in phase space respectively.

Intuitively, we expect this random walk in the space of (local) minima to be
able to explore all states if the ground states form a single cluster, or get
stuck in some part of the manifold if there are several disconnected clusters.
We emphasise that this is slightly different to the dynamical freezing
transition, as that was governed by the size of the energy barriers becoming
larger than the available energy at low temperatures, whereas here we do not
restrict the maximally applied field to induce a transition.


\begin{figure}
  \includegraphics[width=.99\columnwidth]{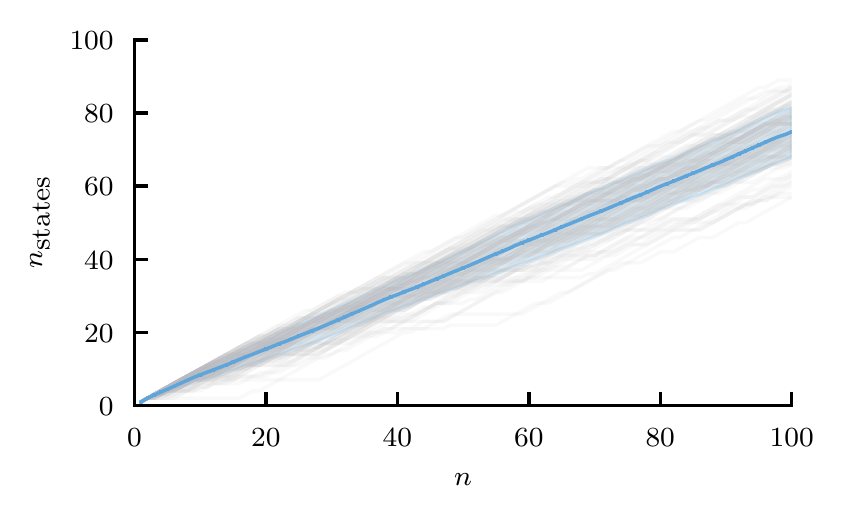}
  \caption{Number of unique encountered states $n_{\textrm{states}}$ when successively forcing the
    state in a random direction until a transition takes place versus number
    of forced transitions $n$. Light-gray lines are individual trajectories for
    different initial states and disorder configurations, shading is the
    standard deviation around the mean.
    \label{fig:random_states}
  }
\end{figure}
In Fig.~\ref{fig:random_states} we show the number of unique encountered states
$n_{\textrm{states}}$ versus the number of forced transitions $n$. Individual
trajectories for different disorder configurations and initial groundstates are
shown as light-gray lines, and the average with standard deviation as the blue
line with shading.

From these results it appears that while individual trajectories can be stuck
for some time in a set of ``known'' states visible in the plateaus, after a few
transitions they do escape and continue to encounter successively more new
states with an increasing number of transitions.

Consequently, the set of ground states appears to be connected in the sense that
successive transitions through thermodynamically small energy barriers allow to
explore a large number of distinct states, e.g. that if they do cluster, that
these clusters are relatively large.

\begin{figure}
  \includegraphics[width=.99\columnwidth]{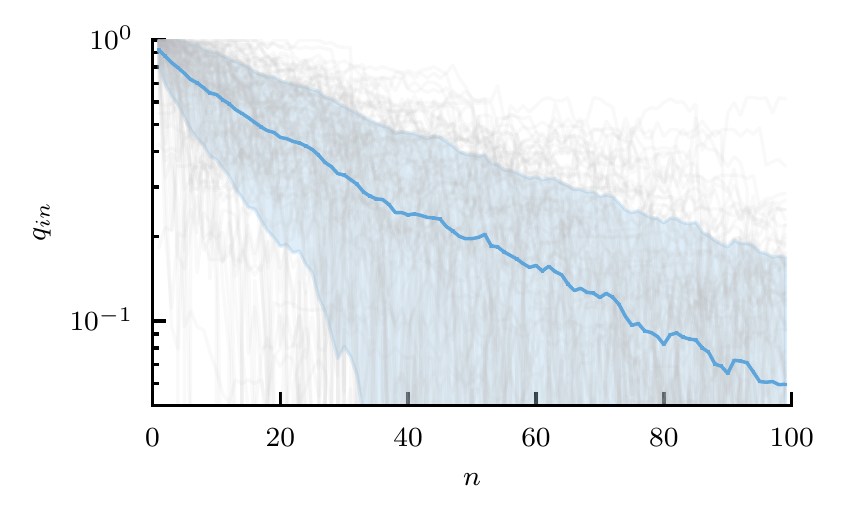}
  \caption{Overlap $q_{in}$ of the initial state with the state reached after
    $n$ forced transitions. Light-gray lines are individual trajectories for
    different initial states and disorder configurations, shading is the
    standard deviation around the mean.
    \label{fig:random_overlap}
  }
\end{figure}
Finally, we consider the overlap $q_{in}$ between the initial state and the
state reached after $n$ transitions in Fig.~\ref{fig:random_overlap}. This is
observed to decay exponentially with the number of forced transitions. Changes
in states appear global, i.e. all spins are changed in every step, but the loss
of overlap is incremental, i.e. we perform a succession of ``small'' steps in an
exponentially large space, ultimately completely decorrelating from the initial
state.

We emphasise that this is very different to the situation expected on the basis
of replica symmetry breaking in a spin glass \cite{Marinari2000} which results
in a hierachical energy landscape
\cite{Parisi1979,Mezard1984,Mezard1986,Charbonneau2014}, and consequently states
stuck in one local basin in which states have a high overlap, separated from
distinct basins by large energy barriers
\cite{Berthier2011,Parisi2006,Charbonneau2014,Pemartin2018}.

Thus, it appears that the set of ground states remains (at least partly)
connected at finite temperatures and can via finite perturbations explore a
large number of states via successive transitions reaching states fully
unrelated to the initial state.

\section{Conclusions\label{sec:conclusions}}
In this work we have studied the dynamics and energy landscape of the recently
identified jammed spin liquid, which has exponentially many exactly degenerate
groundstates in presence of disordered bond couplings.

Since these states are rigid, e.g. they have no zero-energy modes besides global
rotations, they form a discrete set, in contrast to previously studied classical
spin liquids with continuous groundstate manifolds with zero-modes. Despite the
rigidity of the spin ground state configurations there still exist a large
number of very soft normal modes, which appear delocalised with a best-fit
fractional exponent $L^{-5/3}$.

In spite of the rigidity of the ground states, we establish a vanishing
spin-stiffness. The spin autocorrelation shows typical classical spin-liquid
behaviour with an exponential decay rate scaling linearly with temperature in
the intermediate low temperature regime and steepening to $T^{1.3}$ at the
lowest temperatures we access. We also find evidence of spin diffusion, and
obtain a spin diffusion constant $D$ that seems to decrease in the low
temperature regime. However, we are not able to resolve whether diffusion
persists down to the lowest temperatures on larger and larger length-scales with
a finite diffusion constant or disappears completely.

Furthermore, the dynamical structure factor has no sharply defined features,
suggesting that there are no sharp spinwaves present in the disordered model,
but shows concentration of spectral weight at low frequencies, which we
attribute to the large number of soft normal modes of the ground states.

Interestingly, we find a transition (on finite systems) between dynamics that is
able to explore (some parts) of phase space and fully decorrelate from the
original state, and a dynamically arrested regime, in which the states mostly
globally rotate and perform small oscillations being stuck close to an initial
state.

This in turn motivated the detailed study of the energy landscape in terms of
response to forcing of the groundstates. We find that energy barriers between
different groundstates vanish with increasing system size, implying that
excitations, due to finite temperature or perturbations, are able to induce
groundstate transitions. However, there appear to be no local rearrangements
between different groundstates, transitions always requiring a global change in
the spin configuration.

We find that the response is ``anisotropic'', and depends on the form of the
applied force: it is ``gapped'' for forcing in a hard direction, defined via the
spectrum of the Hessian, whereas in a ``soft'' or random direction it smoothly
evolves as a function of the forcing strength. Moreover, local forcing
encounters significantly higher energy barriers and requires larger forcing
fields, which we attribute to the fact that groundstate configurations are
locally rigid and, thus, resist deformation.

Finally, a random walk via successive transitions is able to fully decorrelate
the resulting spin configuration from the initial state in an exponential
fashion, suggesting that the set of ground states remains (at least partly)
connected, and that clusters, if they exist, contain a large number of quite
distinct ground states.

Summarising, we find a complex energy landscape with exponentially many
degenerate discrete locally rigid ground states in a bond-disordered frustrated
magnet, which at finite temperatures or energy densities appear to be connected
within Landau-Lifshitz spin dynamics, with exponentially decaying spin auto
correlations and no sharply defined features in the dynamical structure factor,
and via applied fields, with vanishing energy barriers between ``neighbouring''
distinct groundstates.

We finish by pointing out avenues for further research and open questions.
Firstly, in the kagome Heisenberg antiferromagnet bond disorder does completely
eliminate all zero-modes of the ground states. In light of the recent
connections of frustrated magnetism to topology via spin-origami in the case of
anisotropic interactions in the kagome HAFM
\cite{Roychowdhury2018a,Roychowdhury2018}, where extensive or sub-extensive
numbers of zero-modes were found protected by topological indices, a study of
the (potential) topological features of the jammed spin liquid, which lifts
these degeneracies completely, might provide further insight into the interplay
of frustration and topology in magnets.

Secondly, the arrested regime, with dynamics observed to be ``stuck'' close to a
state interrupted by sudden transitions between distinct states, is reminiscent
of the behaviour in Heisenberg spin glasses \cite{Baity-Jesi2015,Parisi2006}.
Whereas this regime vanishes in the thermodynamic limit in this model, the
presence of this ``glassy'' phase in a model with exactly degenerate states on
finite systems might provide further insight into mechanisms of spin-freezing
and spin glasses.

\sectionn{Acknowledgments} This  work  was  in  part  supported  by  Deutsche
Forschungsgemeinschaft via SFB 1143.


\begin{thebibliography}{47}%
\makeatletter
\providecommand \@ifxundefined [1]{%
 \@ifx{#1\undefined}
}%
\providecommand \@ifnum [1]{%
 \ifnum #1\expandafter \@firstoftwo
 \else \expandafter \@secondoftwo
 \fi
}%
\providecommand \@ifx [1]{%
 \ifx #1\expandafter \@firstoftwo
 \else \expandafter \@secondoftwo
 \fi
}%
\providecommand \natexlab [1]{#1}%
\providecommand \enquote  [1]{``#1''}%
\providecommand \bibnamefont  [1]{#1}%
\providecommand \bibfnamefont [1]{#1}%
\providecommand \citenamefont [1]{#1}%
\providecommand \href@noop [0]{\@secondoftwo}%
\providecommand \href [0]{\begingroup \@sanitize@url \@href}%
\providecommand \@href[1]{\@@startlink{#1}\@@href}%
\providecommand \@@href[1]{\endgroup#1\@@endlink}%
\providecommand \@sanitize@url [0]{\catcode `\\12\catcode `\$12\catcode
  `\&12\catcode `\#12\catcode `\^12\catcode `\_12\catcode `\%12\relax}%
\providecommand \@@startlink[1]{}%
\providecommand \@@endlink[0]{}%
\providecommand \url  [0]{\begingroup\@sanitize@url \@url }%
\providecommand \@url [1]{\endgroup\@href {#1}{\urlprefix }}%
\providecommand \urlprefix  [0]{URL }%
\providecommand \Eprint [0]{\href }%
\providecommand \doibase [0]{http://dx.doi.org/}%
\providecommand \selectlanguage [0]{\@gobble}%
\providecommand \bibinfo  [0]{\@secondoftwo}%
\providecommand \bibfield  [0]{\@secondoftwo}%
\providecommand \translation [1]{[#1]}%
\providecommand \BibitemOpen [0]{}%
\providecommand \bibitemStop [0]{}%
\providecommand \bibitemNoStop [0]{.\EOS\space}%
\providecommand \EOS [0]{\spacefactor3000\relax}%
\providecommand \BibitemShut  [1]{\csname bibitem#1\endcsname}%
\let\auto@bib@innerbib\@empty
\bibitem [{\citenamefont {Reidys}\ and\ \citenamefont
  {Stadler}(2002)}]{Reidys2002}%
  \BibitemOpen
  \bibfield  {author} {\bibinfo {author} {\bibfnamefont {C.~M.}\ \bibnamefont
  {Reidys}}\ and\ \bibinfo {author} {\bibfnamefont {P.~F.}\ \bibnamefont
  {Stadler}},\ }\bibfield  {title} {\enquote {\bibinfo {title} {Combinatorial
  landscapes},}\ }\href {\doibase 10.1137/s0036144501395952} {\bibfield
  {journal} {\bibinfo  {journal} {{SIAM} Review}\ }\textbf {\bibinfo {volume}
  {44}},\ \bibinfo {pages} {3--54} (\bibinfo {year} {2002})}\BibitemShut
  {NoStop}%
\bibitem [{\citenamefont {Franz}\ \emph {et~al.}(2017)\citenamefont {Franz},
  \citenamefont {Parisi}, \citenamefont {Sevelev}, \citenamefont {Urbani},\
  and\ \citenamefont {Zamponi}}]{Franz2017}%
  \BibitemOpen
  \bibfield  {author} {\bibinfo {author} {\bibfnamefont {S.}~\bibnamefont
  {Franz}}, \bibinfo {author} {\bibfnamefont {G.}~\bibnamefont {Parisi}},
  \bibinfo {author} {\bibfnamefont {M.}~\bibnamefont {Sevelev}}, \bibinfo
  {author} {\bibfnamefont {P.}~\bibnamefont {Urbani}}, \ and\ \bibinfo {author}
  {\bibfnamefont {F.}~\bibnamefont {Zamponi}},\ }\bibfield  {title} {\enquote
  {\bibinfo {title} {Universality of the sat-unsat (jamming) threshold in
  non-convex continuous constraint satisfaction problems},}\ }\href {\doibase
  10.21468/SciPostPhys.2.3.019} {\bibfield  {journal} {\bibinfo  {journal}
  {SciPost Phys.}\ }\textbf {\bibinfo {volume} {2}},\ \bibinfo {pages} {019}
  (\bibinfo {year} {2017})}\BibitemShut {NoStop}%
\bibitem [{\citenamefont {{Edwards}}\ and\ \citenamefont
  {{Anderson}}(1975)}]{Anderson1975}%
  \BibitemOpen
  \bibfield  {author} {\bibinfo {author} {\bibfnamefont {S.~F.}\ \bibnamefont
  {{Edwards}}}\ and\ \bibinfo {author} {\bibfnamefont {P.~W.}\ \bibnamefont
  {{Anderson}}},\ }\bibfield  {title} {\enquote {\bibinfo {title} {{Theory of
  spin glasses}},}\ }\href {\doibase 10.1088/0305-4608/5/5/017} {\bibfield
  {journal} {\bibinfo  {journal} {Journal of Physics F Metal Physics}\ }\textbf
  {\bibinfo {volume} {5}},\ \bibinfo {pages} {965--974} (\bibinfo {year}
  {1975})}\BibitemShut {NoStop}%
\bibitem [{\citenamefont {M\'ezard}\ \emph {et~al.}(1984)\citenamefont
  {M\'ezard}, \citenamefont {Parisi}, \citenamefont {Sourlas}, \citenamefont
  {Toulouse},\ and\ \citenamefont {Virasoro}}]{Mezard1984}%
  \BibitemOpen
  \bibfield  {author} {\bibinfo {author} {\bibfnamefont {M.}~\bibnamefont
  {M\'ezard}}, \bibinfo {author} {\bibfnamefont {G.}~\bibnamefont {Parisi}},
  \bibinfo {author} {\bibfnamefont {N.}~\bibnamefont {Sourlas}}, \bibinfo
  {author} {\bibfnamefont {G.}~\bibnamefont {Toulouse}}, \ and\ \bibinfo
  {author} {\bibfnamefont {M.}~\bibnamefont {Virasoro}},\ }\bibfield  {title}
  {\enquote {\bibinfo {title} {Nature of the spin-glass phase},}\ }\href
  {\doibase 10.1103/PhysRevLett.52.1156} {\bibfield  {journal} {\bibinfo
  {journal} {Phys. Rev. Lett.}\ }\textbf {\bibinfo {volume} {52}},\ \bibinfo
  {pages} {1156--1159} (\bibinfo {year} {1984})}\BibitemShut {NoStop}%
\bibitem [{\citenamefont {Mezard}\ \emph {et~al.}(1986)\citenamefont {Mezard},
  \citenamefont {Parisi},\ and\ \citenamefont {Virasoro}}]{Mezard1986}%
  \BibitemOpen
  \bibfield  {author} {\bibinfo {author} {\bibfnamefont {M.}~\bibnamefont
  {Mezard}}, \bibinfo {author} {\bibfnamefont {G.}~\bibnamefont {Parisi}}, \
  and\ \bibinfo {author} {\bibfnamefont {M.}~\bibnamefont {Virasoro}},\ }\href
  {\doibase 10.1142/0271} {\emph {\bibinfo {title} {Spin Glass Theory and
  Beyond}}}\ (\bibinfo  {publisher} {{WORLD} {SCIENTIFIC}},\ \bibinfo {year}
  {1986})\BibitemShut {NoStop}%
\bibitem [{\citenamefont {Young}(1997)}]{Young1997}%
  \BibitemOpen
  \bibfield  {author} {\bibinfo {author} {\bibfnamefont {A.~P.}\ \bibnamefont
  {Young}},\ }\href {\doibase 10.1142/3517} {\emph {\bibinfo {title} {Spin
  Glasses and Random Fields}}}\ (\bibinfo  {publisher} {{WORLD} {SCIENTIFIC}},\
  \bibinfo {year} {1997})\BibitemShut {NoStop}%
\bibitem [{\citenamefont {Charbonneau}\ \emph {et~al.}(2014)\citenamefont
  {Charbonneau}, \citenamefont {Kurchan}, \citenamefont {Parisi}, \citenamefont
  {Urbani},\ and\ \citenamefont {Zamponi}}]{Charbonneau2014}%
  \BibitemOpen
  \bibfield  {author} {\bibinfo {author} {\bibfnamefont {P.}~\bibnamefont
  {Charbonneau}}, \bibinfo {author} {\bibfnamefont {J.}~\bibnamefont
  {Kurchan}}, \bibinfo {author} {\bibfnamefont {G.}~\bibnamefont {Parisi}},
  \bibinfo {author} {\bibfnamefont {P.}~\bibnamefont {Urbani}}, \ and\ \bibinfo
  {author} {\bibfnamefont {F.}~\bibnamefont {Zamponi}},\ }\bibfield  {title}
  {\enquote {\bibinfo {title} {Fractal free energy landscapes in structural
  glasses},}\ }\href {\doibase 10.1038/ncomms4725} {\bibfield  {journal}
  {\bibinfo  {journal} {Nature Communications}\ }\textbf {\bibinfo {volume}
  {5}},\ \bibinfo {pages} {3725} (\bibinfo {year} {2014})}\BibitemShut
  {NoStop}%
\bibitem [{\citenamefont {Charbonneau}\ \emph {et~al.}(2017)\citenamefont
  {Charbonneau}, \citenamefont {Kurchan}, \citenamefont {Parisi}, \citenamefont
  {Urbani},\ and\ \citenamefont {Zamponi}}]{Charbonneau2017}%
  \BibitemOpen
  \bibfield  {author} {\bibinfo {author} {\bibfnamefont {P.}~\bibnamefont
  {Charbonneau}}, \bibinfo {author} {\bibfnamefont {J.}~\bibnamefont
  {Kurchan}}, \bibinfo {author} {\bibfnamefont {G.}~\bibnamefont {Parisi}},
  \bibinfo {author} {\bibfnamefont {P.}~\bibnamefont {Urbani}}, \ and\ \bibinfo
  {author} {\bibfnamefont {F.}~\bibnamefont {Zamponi}},\ }\bibfield  {title}
  {\enquote {\bibinfo {title} {Glass and jamming transitions: From exact
  results to finite-dimensional descriptions},}\ }\href {\doibase
  10.1146/annurev-conmatphys-031016-025334} {\bibfield  {journal} {\bibinfo
  {journal} {Annual Review of Condensed Matter Physics}\ }\textbf {\bibinfo
  {volume} {8}},\ \bibinfo {pages} {265--288} (\bibinfo {year}
  {2017})}\BibitemShut {NoStop}%
\bibitem [{\citenamefont {{Gonzaled-Adalid Pemartin}}\ \emph
  {et~al.}()\citenamefont {{Gonzaled-Adalid Pemartin}}, \citenamefont
  {{Martin-Mayor}}, \citenamefont {{Parisi}},\ and\ \citenamefont
  {{Ruiz-Lorenzo}}}]{Pemartin2018}%
  \BibitemOpen
  \bibfield  {author} {\bibinfo {author} {\bibfnamefont {I.}~\bibnamefont
  {{Gonzaled-Adalid Pemartin}}}, \bibinfo {author} {\bibfnamefont
  {V.}~\bibnamefont {{Martin-Mayor}}}, \bibinfo {author} {\bibfnamefont
  {G.}~\bibnamefont {{Parisi}}}, \ and\ \bibinfo {author} {\bibfnamefont
  {J.~J.}\ \bibnamefont {{Ruiz-Lorenzo}}},\ }\bibfield  {title} {\enquote
  {\bibinfo {title} {{Numerical study of barriers and valleys in the
  free-energy landscape of spin glasses}},}\ }\href@noop {} {\ }\Eprint
  {http://arxiv.org/abs/1811.03414} {arXiv:1811.03414} \BibitemShut {NoStop}%
\bibitem [{\citenamefont {Liu}\ and\ \citenamefont {Nagel}(2010)}]{Liu2010}%
  \BibitemOpen
  \bibfield  {author} {\bibinfo {author} {\bibfnamefont {A.~J.}\ \bibnamefont
  {Liu}}\ and\ \bibinfo {author} {\bibfnamefont {S.~R.}\ \bibnamefont
  {Nagel}},\ }\bibfield  {title} {\enquote {\bibinfo {title} {The jamming
  transition and the marginally jammed solid},}\ }\href {\doibase
  10.1146/annurev-conmatphys-070909-104045} {\bibfield  {journal} {\bibinfo
  {journal} {Annual Review of Condensed Matter Physics}\ }\textbf {\bibinfo
  {volume} {1}},\ \bibinfo {pages} {347--369} (\bibinfo {year}
  {2010})}\BibitemShut {NoStop}%
\bibitem [{\citenamefont {Liu}\ and\ \citenamefont {Nagel}(1998)}]{Liu1998}%
  \BibitemOpen
  \bibfield  {author} {\bibinfo {author} {\bibfnamefont {A.~J.}\ \bibnamefont
  {Liu}}\ and\ \bibinfo {author} {\bibfnamefont {S.~R.}\ \bibnamefont
  {Nagel}},\ }\bibfield  {title} {\enquote {\bibinfo {title} {Nonlinear
  dynamics: Jamming is not just cool any more},}\ }\href {\doibase
  10.1038/23819} {\bibfield  {journal} {\bibinfo  {journal} {Nature}\ }\textbf
  {\bibinfo {volume} {396}},\ \bibinfo {pages} {21--22} (\bibinfo {year}
  {1998})}\BibitemShut {NoStop}%
\bibitem [{\citenamefont {O'Hern}\ \emph {et~al.}(2003)\citenamefont {O'Hern},
  \citenamefont {Silbert}, \citenamefont {Liu},\ and\ \citenamefont
  {Nagel}}]{Ohern2003}%
  \BibitemOpen
  \bibfield  {author} {\bibinfo {author} {\bibfnamefont {C.~S.}\ \bibnamefont
  {O'Hern}}, \bibinfo {author} {\bibfnamefont {L.~E.}\ \bibnamefont {Silbert}},
  \bibinfo {author} {\bibfnamefont {A.~J.}\ \bibnamefont {Liu}}, \ and\
  \bibinfo {author} {\bibfnamefont {S.~R.}\ \bibnamefont {Nagel}},\ }\bibfield
  {title} {\enquote {\bibinfo {title} {Jamming at zero temperature and zero
  applied stress: The epitome of disorder},}\ }\href {\doibase
  10.1103/PhysRevE.68.011306} {\bibfield  {journal} {\bibinfo  {journal} {Phys.
  Rev. E}\ }\textbf {\bibinfo {volume} {68}},\ \bibinfo {pages} {011306}
  (\bibinfo {year} {2003})}\BibitemShut {NoStop}%
\bibitem [{\citenamefont {Berthier}\ and\ \citenamefont
  {Biroli}(2011)}]{Berthier2011}%
  \BibitemOpen
  \bibfield  {author} {\bibinfo {author} {\bibfnamefont {L.}~\bibnamefont
  {Berthier}}\ and\ \bibinfo {author} {\bibfnamefont {G.}~\bibnamefont
  {Biroli}},\ }\bibfield  {title} {\enquote {\bibinfo {title} {Theoretical
  perspective on the glass transition and amorphous materials},}\ }\href
  {\doibase 10.1103/revmodphys.83.587} {\bibfield  {journal} {\bibinfo
  {journal} {Reviews of Modern Physics}\ }\textbf {\bibinfo {volume} {83}},\
  \bibinfo {pages} {587--645} (\bibinfo {year} {2011})}\BibitemShut {NoStop}%
\bibitem [{\citenamefont {Behringer}\ and\ \citenamefont
  {Chakraborty}(2019)}]{Behringer2018}%
  \BibitemOpen
  \bibfield  {author} {\bibinfo {author} {\bibfnamefont {R.~P.}\ \bibnamefont
  {Behringer}}\ and\ \bibinfo {author} {\bibfnamefont {B.}~\bibnamefont
  {Chakraborty}},\ }\bibfield  {title} {\enquote {\bibinfo {title} {The physics
  of jamming for granular materials: a review},}\ }\href
  {http://stacks.iop.org/0034-4885/82/i=1/a=012601} {\bibfield  {journal}
  {\bibinfo  {journal} {Reports on Progress in Physics}\ }\textbf {\bibinfo
  {volume} {82}},\ \bibinfo {pages} {012601} (\bibinfo {year}
  {2019})}\BibitemShut {NoStop}%
\bibitem [{\citenamefont {Onuchic}\ \emph {et~al.}(2000)\citenamefont
  {Onuchic}, \citenamefont {Nymeyer}, \citenamefont {Garc{\'{\i}}a},
  \citenamefont {Chahine},\ and\ \citenamefont {Socci}}]{Onuchic2000}%
  \BibitemOpen
  \bibfield  {author} {\bibinfo {author} {\bibfnamefont {J.~N.}\ \bibnamefont
  {Onuchic}}, \bibinfo {author} {\bibfnamefont {H.}~\bibnamefont {Nymeyer}},
  \bibinfo {author} {\bibfnamefont {A.~E.}\ \bibnamefont {Garc{\'{\i}}a}},
  \bibinfo {author} {\bibfnamefont {J.}~\bibnamefont {Chahine}}, \ and\
  \bibinfo {author} {\bibfnamefont {N.~D.}\ \bibnamefont {Socci}},\ }\bibfield
  {title} {\enquote {\bibinfo {title} {The energy landscape theory of protein
  folding: Insights into folding mechanisms and scenarios},}\ }in\ \href
  {\doibase 10.1016/s0065-3233(00)53003-4} {\emph {\bibinfo {booktitle}
  {Advances in Protein Chemistry}}}\ (\bibinfo  {publisher} {Elsevier},\
  \bibinfo {year} {2000})\ pp.\ \bibinfo {pages} {87--152}\BibitemShut
  {NoStop}%
\bibitem [{\citenamefont {Heidrich}\ \emph {et~al.}(1991)\citenamefont
  {Heidrich}, \citenamefont {Kliesch},\ and\ \citenamefont
  {Quapp}}]{Heidrich1991}%
  \BibitemOpen
  \bibfield  {author} {\bibinfo {author} {\bibfnamefont {D.}~\bibnamefont
  {Heidrich}}, \bibinfo {author} {\bibfnamefont {W.}~\bibnamefont {Kliesch}}, \
  and\ \bibinfo {author} {\bibfnamefont {W.}~\bibnamefont {Quapp}},\ }\href
  {\doibase 10.1007/978-3-642-93499-5} {\emph {\bibinfo {title} {Properties of
  Chemically Interesting Potential Energy Surfaces}}}\ (\bibinfo  {publisher}
  {Springer Berlin Heidelberg},\ \bibinfo {year} {1991})\BibitemShut {NoStop}%
\bibitem [{\citenamefont {Wright}(1932)}]{Wright1932}%
  \BibitemOpen
  \bibfield  {author} {\bibinfo {author} {\bibfnamefont {S.}~\bibnamefont
  {Wright}},\ }\bibfield  {title} {\enquote {\bibinfo {title} {{The roles of
  mutation, inbreeding, crossbreeding and selection in evolution}},}\
  }\href@noop {} {\bibfield  {journal} {\bibinfo  {journal} {Proceedings of the
  Sixth International Congress of Genetics}\ }\textbf {\bibinfo {volume} {1}},\
  \bibinfo {pages} {356--366} (\bibinfo {year} {1932})}\BibitemShut {NoStop}%
\bibitem [{\citenamefont {Mustonen}\ and\ \citenamefont
  {Lässig}(2009)}]{Mustonen2009}%
  \BibitemOpen
  \bibfield  {author} {\bibinfo {author} {\bibfnamefont {V.}~\bibnamefont
  {Mustonen}}\ and\ \bibinfo {author} {\bibfnamefont {M.}~\bibnamefont
  {Lässig}},\ }\bibfield  {title} {\enquote {\bibinfo {title} {From fitness
  landscapes to seascapes: non-equilibrium dynamics of selection and
  adaptation},}\ }\href {\doibase 10.1016/j.tig.2009.01.002} {\bibfield
  {journal} {\bibinfo  {journal} {Trends in Genetics}\ }\textbf {\bibinfo
  {volume} {25}},\ \bibinfo {pages} {111--119} (\bibinfo {year}
  {2009})}\BibitemShut {NoStop}%
\bibitem [{\citenamefont {Stadler}(2002)}]{Stadler2002}%
  \BibitemOpen
  \bibfield  {author} {\bibinfo {author} {\bibfnamefont {P.~F.}\ \bibnamefont
  {Stadler}},\ }\enquote {\bibinfo {title} {Fitness landscapes},}\ in\ \href
  {\doibase 10.1007/3-540-45692-9_10} {\emph {\bibinfo {booktitle} {Biological
  Evolution and Statistical Physics}}},\ \bibinfo {editor} {edited by\ \bibinfo
  {editor} {\bibfnamefont {M.}~\bibnamefont {L{\"a}ssig}}\ and\ \bibinfo
  {editor} {\bibfnamefont {A.}~\bibnamefont {Valleriani}}}\ (\bibinfo
  {publisher} {Springer Berlin Heidelberg},\ \bibinfo {address} {Berlin,
  Heidelberg},\ \bibinfo {year} {2002})\ pp.\ \bibinfo {pages}
  {183--204}\BibitemShut {NoStop}%
\bibitem [{\citenamefont {Hartl}(2014)}]{Hartl2014}%
  \BibitemOpen
  \bibfield  {author} {\bibinfo {author} {\bibfnamefont {D.~L.}\ \bibnamefont
  {Hartl}},\ }\bibfield  {title} {\enquote {\bibinfo {title} {What can we learn
  from fitness landscapes?}}\ }\href {\doibase 10.1016/j.mib.2014.08.001}
  {\bibfield  {journal} {\bibinfo  {journal} {Current Opinion in Microbiology}\
  }\textbf {\bibinfo {volume} {21}},\ \bibinfo {pages} {51--57} (\bibinfo
  {year} {2014})}\BibitemShut {NoStop}%
\bibitem [{\citenamefont {Barahona}(1982)}]{Barahona1982}%
  \BibitemOpen
  \bibfield  {author} {\bibinfo {author} {\bibfnamefont {F.}~\bibnamefont
  {Barahona}},\ }\bibfield  {title} {\enquote {\bibinfo {title} {On the
  computational complexity of ising spin glass models},}\ }\href
  {http://stacks.iop.org/0305-4470/15/i=10/a=028} {\bibfield  {journal}
  {\bibinfo  {journal} {Journal of Physics A: Mathematical and General}\
  }\textbf {\bibinfo {volume} {15}},\ \bibinfo {pages} {3241} (\bibinfo {year}
  {1982})}\BibitemShut {NoStop}%
\bibitem [{\citenamefont {Anderson}(1956)}]{Anderson1956}%
  \BibitemOpen
  \bibfield  {author} {\bibinfo {author} {\bibfnamefont {P.~W.}\ \bibnamefont
  {Anderson}},\ }\bibfield  {title} {\enquote {\bibinfo {title} {Ordering and
  antiferromagnetism in ferrites},}\ }\href {\doibase 10.1103/PhysRev.102.1008}
  {\bibfield  {journal} {\bibinfo  {journal} {Phys. Rev.}\ }\textbf {\bibinfo
  {volume} {102}},\ \bibinfo {pages} {1008--1013} (\bibinfo {year}
  {1956})}\BibitemShut {NoStop}%
\bibitem [{\citenamefont {Villain}(1979)}]{Villain1979}%
  \BibitemOpen
  \bibfield  {author} {\bibinfo {author} {\bibfnamefont {J.}~\bibnamefont
  {Villain}},\ }\bibfield  {title} {\enquote {\bibinfo {title} {Insulating spin
  glasses},}\ }\href {\doibase 10.1007/BF01325811} {\bibfield  {journal}
  {\bibinfo  {journal} {Zeitschrift f{\"u}r Physik B Condensed Matter}\
  }\textbf {\bibinfo {volume} {33}},\ \bibinfo {pages} {31--42} (\bibinfo
  {year} {1979})}\BibitemShut {NoStop}%
\bibitem [{\citenamefont {Chalker}\ \emph {et~al.}(1992)\citenamefont
  {Chalker}, \citenamefont {Holdsworth},\ and\ \citenamefont
  {Shender}}]{Chalker1992}%
  \BibitemOpen
  \bibfield  {author} {\bibinfo {author} {\bibfnamefont {J.~T.}\ \bibnamefont
  {Chalker}}, \bibinfo {author} {\bibfnamefont {P.~C.~W.}\ \bibnamefont
  {Holdsworth}}, \ and\ \bibinfo {author} {\bibfnamefont {E.~F.}\ \bibnamefont
  {Shender}},\ }\bibfield  {title} {\enquote {\bibinfo {title} {Hidden order in
  a frustrated system: Properties of the heisenberg kagom\'e
  antiferromagnet},}\ }\href {\doibase 10.1103/PhysRevLett.68.855} {\bibfield
  {journal} {\bibinfo  {journal} {Phys. Rev. Lett.}\ }\textbf {\bibinfo
  {volume} {68}},\ \bibinfo {pages} {855--858} (\bibinfo {year}
  {1992})}\BibitemShut {NoStop}%
\bibitem [{\citenamefont {Moessner}\ and\ \citenamefont
  {Chalker}(1998{\natexlab{a}})}]{Moessner1998}%
  \BibitemOpen
  \bibfield  {author} {\bibinfo {author} {\bibfnamefont {R.}~\bibnamefont
  {Moessner}}\ and\ \bibinfo {author} {\bibfnamefont {J.~T.}\ \bibnamefont
  {Chalker}},\ }\bibfield  {title} {\enquote {\bibinfo {title} {Low-temperature
  properties of classical geometrically frustrated antiferromagnets},}\ }\href
  {\doibase 10.1103/PhysRevB.58.12049} {\bibfield  {journal} {\bibinfo
  {journal} {Phys. Rev. B}\ }\textbf {\bibinfo {volume} {58}},\ \bibinfo
  {pages} {12049--12062} (\bibinfo {year} {1998}{\natexlab{a}})}\BibitemShut
  {NoStop}%
\bibitem [{\citenamefont {Huse}\ and\ \citenamefont
  {Rutenberg}(1992)}]{Huse1992}%
  \BibitemOpen
  \bibfield  {author} {\bibinfo {author} {\bibfnamefont {D.~A.}\ \bibnamefont
  {Huse}}\ and\ \bibinfo {author} {\bibfnamefont {A.~D.}\ \bibnamefont
  {Rutenberg}},\ }\bibfield  {title} {\enquote {\bibinfo {title} {Classical
  antiferromagnets on the kagom\'e lattice},}\ }\href {\doibase
  10.1103/PhysRevB.45.7536} {\bibfield  {journal} {\bibinfo  {journal} {Phys.
  Rev. B}\ }\textbf {\bibinfo {volume} {45}},\ \bibinfo {pages} {7536--7539}
  (\bibinfo {year} {1992})}\BibitemShut {NoStop}%
\bibitem [{\citenamefont {Reimers}\ and\ \citenamefont
  {Berlinsky}(1993)}]{Reimers1993}%
  \BibitemOpen
  \bibfield  {author} {\bibinfo {author} {\bibfnamefont {J.~N.}\ \bibnamefont
  {Reimers}}\ and\ \bibinfo {author} {\bibfnamefont {A.~J.}\ \bibnamefont
  {Berlinsky}},\ }\bibfield  {title} {\enquote {\bibinfo {title} {Order by
  disorder in the classical heisenberg kagom\'e antiferromagnet},}\ }\href
  {\doibase 10.1103/PhysRevB.48.9539} {\bibfield  {journal} {\bibinfo
  {journal} {Phys. Rev. B}\ }\textbf {\bibinfo {volume} {48}},\ \bibinfo
  {pages} {9539--9554} (\bibinfo {year} {1993})}\BibitemShut {NoStop}%
\bibitem [{\citenamefont {Zhitomirsky}(2008)}]{Zhitomirsky2008}%
  \BibitemOpen
  \bibfield  {author} {\bibinfo {author} {\bibfnamefont {M.~E.}\ \bibnamefont
  {Zhitomirsky}},\ }\bibfield  {title} {\enquote {\bibinfo {title} {Octupolar
  ordering of classical kagome antiferromagnets in two and three dimensions},}\
  }\href {\doibase 10.1103/PhysRevB.78.094423} {\bibfield  {journal} {\bibinfo
  {journal} {Phys. Rev. B}\ }\textbf {\bibinfo {volume} {78}},\ \bibinfo
  {pages} {094423} (\bibinfo {year} {2008})}\BibitemShut {NoStop}%
\bibitem [{\citenamefont {Chern}\ and\ \citenamefont
  {Moessner}(2013)}]{Chern2013}%
  \BibitemOpen
  \bibfield  {author} {\bibinfo {author} {\bibfnamefont {G.-W.}\ \bibnamefont
  {Chern}}\ and\ \bibinfo {author} {\bibfnamefont {R.}~\bibnamefont
  {Moessner}},\ }\bibfield  {title} {\enquote {\bibinfo {title} {Dipolar order
  by disorder in the classical heisenberg antiferromagnet on the kagome
  lattice},}\ }\href {\doibase 10.1103/PhysRevLett.110.077201} {\bibfield
  {journal} {\bibinfo  {journal} {Phys. Rev. Lett.}\ }\textbf {\bibinfo
  {volume} {110}},\ \bibinfo {pages} {077201} (\bibinfo {year}
  {2013})}\BibitemShut {NoStop}%
\bibitem [{\citenamefont {Roychowdhury}\ and\ \citenamefont
  {Lawler}(2018)}]{Roychowdhury2018a}%
  \BibitemOpen
  \bibfield  {author} {\bibinfo {author} {\bibfnamefont {K.}~\bibnamefont
  {Roychowdhury}}\ and\ \bibinfo {author} {\bibfnamefont {M.~J.}\ \bibnamefont
  {Lawler}},\ }\bibfield  {title} {\enquote {\bibinfo {title} {Classification
  of magnetic frustration and metamaterials from topology},}\ }\href {\doibase
  10.1103/PhysRevB.98.094432} {\bibfield  {journal} {\bibinfo  {journal} {Phys.
  Rev. B}\ }\textbf {\bibinfo {volume} {98}},\ \bibinfo {pages} {094432}
  (\bibinfo {year} {2018})}\BibitemShut {NoStop}%
\bibitem [{\citenamefont {Roychowdhury}\ \emph {et~al.}(2018)\citenamefont
  {Roychowdhury}, \citenamefont {Rocklin},\ and\ \citenamefont
  {Lawler}}]{Roychowdhury2018}%
  \BibitemOpen
  \bibfield  {author} {\bibinfo {author} {\bibfnamefont {K.}~\bibnamefont
  {Roychowdhury}}, \bibinfo {author} {\bibfnamefont {D.~Z.}\ \bibnamefont
  {Rocklin}}, \ and\ \bibinfo {author} {\bibfnamefont {M.~J.}\ \bibnamefont
  {Lawler}},\ }\bibfield  {title} {\enquote {\bibinfo {title} {Topology and
  geometry of spin origami},}\ }\href {\doibase 10.1103/PhysRevLett.121.177201}
  {\bibfield  {journal} {\bibinfo  {journal} {Phys. Rev. Lett.}\ }\textbf
  {\bibinfo {volume} {121}},\ \bibinfo {pages} {177201} (\bibinfo {year}
  {2018})}\BibitemShut {NoStop}%
\bibitem [{\citenamefont {Bilitewski}\ \emph {et~al.}(2017)\citenamefont
  {Bilitewski}, \citenamefont {Zhitomirsky},\ and\ \citenamefont
  {Moessner}}]{Bilitewski2017}%
  \BibitemOpen
  \bibfield  {author} {\bibinfo {author} {\bibfnamefont {T.}~\bibnamefont
  {Bilitewski}}, \bibinfo {author} {\bibfnamefont {M.~E.}\ \bibnamefont
  {Zhitomirsky}}, \ and\ \bibinfo {author} {\bibfnamefont {R.}~\bibnamefont
  {Moessner}},\ }\bibfield  {title} {\enquote {\bibinfo {title} {{Jammed Spin
  Liquid in the Bond-Disordered Kagome Antiferromagnet}},}\ }\href {\doibase
  10.1103/PhysRevLett.119.247201} {\bibfield  {journal} {\bibinfo  {journal}
  {Phys. Rev. Lett.}\ }\textbf {\bibinfo {volume} {119}},\ \bibinfo {pages}
  {247201} (\bibinfo {year} {2017})}\BibitemShut {NoStop}%
\bibitem [{\citenamefont {L.~D.~Landau}(1975)}]{Landau1975}%
  \BibitemOpen
  \bibfield  {author} {\bibinfo {author} {\bibfnamefont {E.~M.~L.}\
  \bibnamefont {L.~D.~Landau}},\ }\href@noop {} {\emph {\bibinfo {title} {The
  Classical Theory of Fields}}},\ \bibinfo {edition} {4th}\ ed.\ (\bibinfo
  {publisher} {Butterworth-Heinemann,Oxford},\ \bibinfo {year}
  {1975})\BibitemShut {NoStop}%
\bibitem [{\citenamefont {Moessner}\ and\ \citenamefont
  {Chalker}(1998{\natexlab{b}})}]{Moessner1998a}%
  \BibitemOpen
  \bibfield  {author} {\bibinfo {author} {\bibfnamefont {R.}~\bibnamefont
  {Moessner}}\ and\ \bibinfo {author} {\bibfnamefont {J.~T.}\ \bibnamefont
  {Chalker}},\ }\bibfield  {title} {\enquote {\bibinfo {title} {Properties of a
  classical spin liquid: The heisenberg pyrochlore antiferromagnet},}\ }\href
  {\doibase 10.1103/PhysRevLett.80.2929} {\bibfield  {journal} {\bibinfo
  {journal} {Phys. Rev. Lett.}\ }\textbf {\bibinfo {volume} {80}},\ \bibinfo
  {pages} {2929--2932} (\bibinfo {year} {1998}{\natexlab{b}})}\BibitemShut
  {NoStop}%
\bibitem [{\citenamefont {Conlon}\ and\ \citenamefont
  {Chalker}(2009)}]{Conlon2009}%
  \BibitemOpen
  \bibfield  {author} {\bibinfo {author} {\bibfnamefont {P.~H.}\ \bibnamefont
  {Conlon}}\ and\ \bibinfo {author} {\bibfnamefont {J.~T.}\ \bibnamefont
  {Chalker}},\ }\bibfield  {title} {\enquote {\bibinfo {title} {{Spin Dynamics
  in Pyrochlore Heisenberg Antiferromagnets}},}\ }\href {\doibase
  10.1103/PhysRevLett.102.237206} {\bibfield  {journal} {\bibinfo  {journal}
  {Phys. Rev. Lett.}\ }\textbf {\bibinfo {volume} {102}},\ \bibinfo {pages}
  {237206} (\bibinfo {year} {2009})}\BibitemShut {NoStop}%
\bibitem [{\citenamefont {Robert}\ \emph {et~al.}(2008)\citenamefont {Robert},
  \citenamefont {Canals}, \citenamefont {Simonet},\ and\ \citenamefont
  {Ballou}}]{Robert2008}%
  \BibitemOpen
  \bibfield  {author} {\bibinfo {author} {\bibfnamefont {J.}~\bibnamefont
  {Robert}}, \bibinfo {author} {\bibfnamefont {B.}~\bibnamefont {Canals}},
  \bibinfo {author} {\bibfnamefont {V.}~\bibnamefont {Simonet}}, \ and\
  \bibinfo {author} {\bibfnamefont {R.}~\bibnamefont {Ballou}},\ }\bibfield
  {title} {\enquote {\bibinfo {title} {Propagation and ghosts in the classical
  kagome antiferromagnet},}\ }\href {\doibase 10.1103/PhysRevLett.101.117207}
  {\bibfield  {journal} {\bibinfo  {journal} {Phys. Rev. Lett.}\ }\textbf
  {\bibinfo {volume} {101}},\ \bibinfo {pages} {117207} (\bibinfo {year}
  {2008})}\BibitemShut {NoStop}%
\bibitem [{\citenamefont {Taillefumier}\ \emph {et~al.}(2014)\citenamefont
  {Taillefumier}, \citenamefont {Robert}, \citenamefont {Henley}, \citenamefont
  {Moessner},\ and\ \citenamefont {Canals}}]{Taillefumier2014}%
  \BibitemOpen
  \bibfield  {author} {\bibinfo {author} {\bibfnamefont {M.}~\bibnamefont
  {Taillefumier}}, \bibinfo {author} {\bibfnamefont {J.}~\bibnamefont
  {Robert}}, \bibinfo {author} {\bibfnamefont {C.~L.}\ \bibnamefont {Henley}},
  \bibinfo {author} {\bibfnamefont {R.}~\bibnamefont {Moessner}}, \ and\
  \bibinfo {author} {\bibfnamefont {B.}~\bibnamefont {Canals}},\ }\bibfield
  {title} {\enquote {\bibinfo {title} {{Semiclassical Spin Dynamics of the
  Antiferromagnetic Heisenberg Model on the Kagome Lattice}},}\ }\href
  {\doibase 10.1103/PhysRevB.90.064419} {\bibfield  {journal} {\bibinfo
  {journal} {Phys. Rev. B}\ }\textbf {\bibinfo {volume} {90}},\ \bibinfo
  {pages} {064419} (\bibinfo {year} {2014})}\BibitemShut {NoStop}%
\bibitem [{\citenamefont {M\"uller}(1988)}]{Mueller1988}%
  \BibitemOpen
  \bibfield  {author} {\bibinfo {author} {\bibfnamefont {G.}~\bibnamefont
  {M\"uller}},\ }\bibfield  {title} {\enquote {\bibinfo {title} {Anomalous spin
  diffusion in classical heisenberg magnets},}\ }\href {\doibase
  10.1103/PhysRevLett.60.2785} {\bibfield  {journal} {\bibinfo  {journal}
  {Phys. Rev. Lett.}\ }\textbf {\bibinfo {volume} {60}},\ \bibinfo {pages}
  {2785--2788} (\bibinfo {year} {1988})}\BibitemShut {NoStop}%
\bibitem [{\citenamefont {Gerling}\ and\ \citenamefont
  {Landau}(1989)}]{Gerling1989}%
  \BibitemOpen
  \bibfield  {author} {\bibinfo {author} {\bibfnamefont {R.~W.}\ \bibnamefont
  {Gerling}}\ and\ \bibinfo {author} {\bibfnamefont {D.~P.}\ \bibnamefont
  {Landau}},\ }\bibfield  {title} {\enquote {\bibinfo {title} {Comment on
  anomalous spin diffusion in classical heisenberg magnets.}}\ }\href {\doibase
  10.1103/PhysRevLett.63.812} {\bibfield  {journal} {\bibinfo  {journal} {Phys.
  Rev. Lett.}\ }\textbf {\bibinfo {volume} {63}},\ \bibinfo {pages} {812--812}
  (\bibinfo {year} {1989})}\BibitemShut {NoStop}%
\bibitem [{\citenamefont {M\"uller}(1989)}]{Mueller1989}%
  \BibitemOpen
  \bibfield  {author} {\bibinfo {author} {\bibfnamefont {G.}~\bibnamefont
  {M\"uller}},\ }\bibfield  {title} {\enquote {\bibinfo {title} {M\"uller
  replies:},}\ }\href {\doibase 10.1103/PhysRevLett.63.813} {\bibfield
  {journal} {\bibinfo  {journal} {Phys. Rev. Lett.}\ }\textbf {\bibinfo
  {volume} {63}},\ \bibinfo {pages} {813--813} (\bibinfo {year}
  {1989})}\BibitemShut {NoStop}%
\bibitem [{\citenamefont {Gerling}\ and\ \citenamefont
  {Landau}(1990)}]{Gerling1990}%
  \BibitemOpen
  \bibfield  {author} {\bibinfo {author} {\bibfnamefont {R.~W.}\ \bibnamefont
  {Gerling}}\ and\ \bibinfo {author} {\bibfnamefont {D.~P.}\ \bibnamefont
  {Landau}},\ }\bibfield  {title} {\enquote {\bibinfo {title} {Time-dependent
  behavior of classical spin chains at infinite temperature},}\ }\href
  {\doibase 10.1103/PhysRevB.42.8214} {\bibfield  {journal} {\bibinfo
  {journal} {Phys. Rev. B}\ }\textbf {\bibinfo {volume} {42}},\ \bibinfo
  {pages} {8214--8219} (\bibinfo {year} {1990})}\BibitemShut {NoStop}%
\bibitem [{\citenamefont {Parisi}(2006)}]{Parisi2006}%
  \BibitemOpen
  \bibfield  {author} {\bibinfo {author} {\bibfnamefont {G.}~\bibnamefont
  {Parisi}},\ }\bibfield  {title} {\enquote {\bibinfo {title} {Spin glasses and
  fragile glasses: Statics, dynamics, and complexity},}\ }\href {\doibase
  10.1073/pnas.0601120103} {\bibfield  {journal} {\bibinfo  {journal}
  {Proceedings of the National Academy of Sciences}\ }\textbf {\bibinfo
  {volume} {103}},\ \bibinfo {pages} {7948--7955} (\bibinfo {year}
  {2006})}\BibitemShut {NoStop}%
\bibitem [{\citenamefont {Baity-Jesi}\ \emph {et~al.}(2015)\citenamefont
  {Baity-Jesi}, \citenamefont {Mart\'{\i}n-Mayor}, \citenamefont {Parisi},\
  and\ \citenamefont {Perez-Gaviro}}]{Baity-Jesi2015}%
  \BibitemOpen
  \bibfield  {author} {\bibinfo {author} {\bibfnamefont {M.}~\bibnamefont
  {Baity-Jesi}}, \bibinfo {author} {\bibfnamefont {V.}~\bibnamefont
  {Mart\'{\i}n-Mayor}}, \bibinfo {author} {\bibfnamefont {G.}~\bibnamefont
  {Parisi}}, \ and\ \bibinfo {author} {\bibfnamefont {S.}~\bibnamefont
  {Perez-Gaviro}},\ }\bibfield  {title} {\enquote {\bibinfo {title} {Soft
  modes, localization, and two-level systems in spin glasses},}\ }\href
  {\doibase 10.1103/PhysRevLett.115.267205} {\bibfield  {journal} {\bibinfo
  {journal} {Phys. Rev. Lett.}\ }\textbf {\bibinfo {volume} {115}},\ \bibinfo
  {pages} {267205} (\bibinfo {year} {2015})}\BibitemShut {NoStop}%
\bibitem [{\citenamefont {Mazzacurati}\ \emph {et~al.}(1996)\citenamefont
  {Mazzacurati}, \citenamefont {Ruocco},\ and\ \citenamefont
  {Sampoli}}]{Mazzacurati1996}%
  \BibitemOpen
  \bibfield  {author} {\bibinfo {author} {\bibfnamefont {V.}~\bibnamefont
  {Mazzacurati}}, \bibinfo {author} {\bibfnamefont {G.}~\bibnamefont {Ruocco}},
  \ and\ \bibinfo {author} {\bibfnamefont {M.}~\bibnamefont {Sampoli}},\
  }\bibfield  {title} {\enquote {\bibinfo {title} {Low-frequency atomic motion
  in a model glass},}\ }\href {http://stacks.iop.org/0295-5075/34/i=9/a=681}
  {\bibfield  {journal} {\bibinfo  {journal} {EPL (Europhysics Letters)}\
  }\textbf {\bibinfo {volume} {34}},\ \bibinfo {pages} {681} (\bibinfo {year}
  {1996})}\BibitemShut {NoStop}%
\bibitem [{\citenamefont {Bell}\ and\ \citenamefont {Dean}(1970)}]{Bell1970}%
  \BibitemOpen
  \bibfield  {author} {\bibinfo {author} {\bibfnamefont {R.~J.}\ \bibnamefont
  {Bell}}\ and\ \bibinfo {author} {\bibfnamefont {P.}~\bibnamefont {Dean}},\
  }\bibfield  {title} {\enquote {\bibinfo {title} {Atomic vibrations in
  vitreous silica},}\ }\href {\doibase 10.1039/df9705000055} {\bibfield
  {journal} {\bibinfo  {journal} {Discussions of the Faraday Society}\ }\textbf
  {\bibinfo {volume} {50}},\ \bibinfo {pages} {55} (\bibinfo {year}
  {1970})}\BibitemShut {NoStop}%
\bibitem [{\citenamefont {Marinari}\ \emph {et~al.}(2000)\citenamefont
  {Marinari}, \citenamefont {Parisi}, \citenamefont {Ricci-Tersenghi},
  \citenamefont {Ruiz-Lorenzo},\ and\ \citenamefont {Zuliani}}]{Marinari2000}%
  \BibitemOpen
  \bibfield  {author} {\bibinfo {author} {\bibfnamefont {E.}~\bibnamefont
  {Marinari}}, \bibinfo {author} {\bibfnamefont {G.}~\bibnamefont {Parisi}},
  \bibinfo {author} {\bibfnamefont {F.}~\bibnamefont {Ricci-Tersenghi}},
  \bibinfo {author} {\bibfnamefont {J.~J.}\ \bibnamefont {Ruiz-Lorenzo}}, \
  and\ \bibinfo {author} {\bibfnamefont {F.}~\bibnamefont {Zuliani}},\
  }\bibfield  {title} {\enquote {\bibinfo {title} {Replica symmetry breaking in
  short-range spin glasses: Theoretical foundations and numerical evidences},}\
  }\href {\doibase 10.1023/A:1018607809852} {\bibfield  {journal} {\bibinfo
  {journal} {Journal of Statistical Physics}\ }\textbf {\bibinfo {volume}
  {98}},\ \bibinfo {pages} {973--1074} (\bibinfo {year} {2000})}\BibitemShut
  {NoStop}%
\bibitem [{\citenamefont {Parisi}(1979)}]{Parisi1979}%
  \BibitemOpen
  \bibfield  {author} {\bibinfo {author} {\bibfnamefont {G.}~\bibnamefont
  {Parisi}},\ }\bibfield  {title} {\enquote {\bibinfo {title} {Infinite number
  of order parameters for spin-glasses},}\ }\href {\doibase
  10.1103/PhysRevLett.43.1754} {\bibfield  {journal} {\bibinfo  {journal}
  {Phys. Rev. Lett.}\ }\textbf {\bibinfo {volume} {43}},\ \bibinfo {pages}
  {1754--1756} (\bibinfo {year} {1979})}\BibitemShut {NoStop}%
\end{thebibliography}

%

\end{document}